%% file: article_aa30797_17.tex
\documentclass[traditabstract]{aa}

\usepackage{graphicx}
\usepackage{txfonts}
\usepackage{color}
\usepackage[switch, modulo]{lineno}
\usepackage{rotating}
\usepackage{xspace}
\usepackage{float}
\usepackage{url}

\usepackage{gensymb}
\usepackage{threeparttable}

\newcommand{\g}{$\gamma$\xspace}
\newcommand{\wco}{$W_{\rm{CO}}$\xspace}
\newcommand{\wcounit}{K km s$^{-1}$\xspace}
\newcommand{\fcodk}{$f_{\rm{COdk \, H_2}}$\xspace}
\newcommand{\fcodkm}{$F_{\rm{COdk \, H_2}}$\xspace}

\newcommand{\hd}{H$_2$\xspace}
\newcommand{\cosat}{$\rm{CO_{\rm sat}}$\xspace}
\newcommand{\hi}{{H\,{\sc i} }\xspace}

\newcommand{\cii}{{[C\,{\sc ii}] }\xspace}
\newcommand{\nh}{$N_{\rm{H}}$\xspace}
\newcommand{\nhi}{$N_{\rm{HI}}$\xspace}

\newcommand{\nhd}{$N_{\rm{H}_2}$\xspace}
\newcommand{\nhdnm}{$N_{\rm{H}}^{\rm{DNM}}$\xspace}
\newcommand{\nhtot}{$N_{\rm{H}}^{\rm{tot}}$\xspace}
\newcommand{\nhdtot}{$N_{\rm{H_2}}^{\rm{tot}}$\xspace}
\newcommand{\opa}{$\tau_{353}/N_{\rm{H}}$\xspace}
\newcommand{\opavghi}{$\overline{\tau_{353}/N}_{\rm{H}}^{\rm{HI}}$\xspace}
\newcommand{\opavgdnm}{$\overline{\tau_{353}/N}_{\rm{H}}^{\rm{DNM}}$\xspace}
\newcommand{\opavgco}{$\overline{\tau_{353}/N}_{\rm{H}}^{\rm{CO}}$\xspace}
\newcommand{\opaunit}{$10^{-27}$ cm$^2$ H$^{-1}$\xspace}
\newcommand{\av}{$A_{\rm{V}}$\xspace}
\newcommand{\ajj}{$A_{\rm{J}}$\xspace}

\newcommand{\xco}{$X_{\rm{CO}}$\xspace}

\newcommand{\xcounit}{$10^{20}$ cm$^{-2}$ K$^{-1}$ km$^{-1}$ s\xspace}
\newcommand{\taunu}{$\tau_{353}$\xspace}

\newcommand{\msq}{$^{-2} $\xspace}

\definecolor{deepPink}{RGB}{255,20,147}
\definecolor{firebrick}{RGB}{178,34,34}
\definecolor{red}{RGB}{255, 0, 0}
\definecolor{darkGrey}{RGB}{172,172,172}
\definecolor{yellow}{RGB}{255, 255, 0}
\definecolor{darkOrange}{RGB}{255 ,145, 0}
\definecolor{dodgerBlue}{RGB}{30, 144, 255}
\definecolor{navy}{RGB}{0, 0 ,128}
\definecolor{lime}{RGB}{0., 255, 0}
\definecolor{mediumOrchid}{RGB}{186,85,211}
\definecolor{goldenRod}{RGB}{228,175,32}
\definecolor{cyan}{RGB}{0, 255 ,255}
\definecolor{aquaMarine}{RGB}{127,255,212}
\definecolor{black}{RGB}{0, 0 ,0}
\definecolor{green}{RGB}{0, 128, 0}

\begin{document}

   \title{Cosmic-rays, gas, and dust in nearby anticentre clouds : \\ II -- Interstellar phase transitions and the Dark Neutral Medium}

\author{
Q.~Remy$^{(1)}$ \and 
I.~A.~Grenier$^{(1)}$ \and 
D.J.~Marshall$^{(1)}$ \and 
J.~M.~Casandjian$^{(1)}$ 
}
\authorrunning{LAT collaboration}

\institute{
\inst{1}~Laboratoire AIM, CEA-IRFU/CNRS/Universit\'e Paris Diderot, Service d'Astrophysique, CEA Saclay, F-91191 Gif sur Yvette, France\\ 
\email{quentin.remy@cea.fr} \\
\email{isabelle.grenier@cea.fr} \\
}

   \date{Received 16 March 2017 / Accepted 9 November 2017}
   
\abstract   




\abstract
{}
{ \hi 21-cm and $^{12}$CO 2.6-mm line emissions respectively trace the atomic and molecular gas phases, but they miss most of the opaque \hi and diffuse \hd present in the Dark Neutral Medium (DNM) at the transition between the \hi-bright and CO-bright regions. Jointly probing {H\,{\sc i}}, CO, and DNM gas, we aim to constrain the threshold of the \hi-\hd transition in visual extinction, \av, and in total hydrogen column densities, \nhtot. We also aim to measure gas mass fractions in the different phases and to test their relation to cloud properties. } 
{We have used dust optical depth measurements at 353 GHz, \g-ray maps at GeV energies, and \hi and CO line data to trace the gas column densities and map the DNM in nearby clouds toward the Galactic anticentre and Chamaeleon regions. We have selected a subset of 15 individual clouds, from diffuse to star-forming structures, in order to study the different phases across each cloud and to probe changes from cloud to cloud.}
{The atomic fraction of the total hydrogen column density is observed to decrease in the (0.6--1) $\times 10^{21}$ cm$^{-2}$ range in \nhtot (\av$\approx0.4$ magnitude) because of the formation of \hd molecules. 
The onset of detectable CO intensities varies by only a factor of 4 from cloud to cloud, between 0.6$ \times 10^{21}$ cm$^{-2}$ and 2.5$ \times 10^{21}$ cm$^{-2}$ in total gas column density. 
We observe larger \hd column densities than linearly inferred from the CO intensities at \av $> 3$ magnitudes because of the large CO optical thickness: the additional \hd mass in this regime represents on average 20$\%$ of the CO-inferred molecular mass. In the DNM envelopes, we find that the fraction of diffuse CO-dark \hd in the molecular column densities  decreases with increasing \av in a cloud. For a half molecular DNM, the fraction decreases from more than 80$\%$ at 0.4 magnitude to less than 20$\%$ beyond 2 magnitudes. In mass, the DNM fraction varies with the cloud properties. Clouds with low peak CO intensities exhibit large CO-dark \hd fractions in molecular mass, in particular the diffuse clouds lying at high altitude above the Galactic plane. The mass present in the DNM envelopes appears to scale with the molecular mass seen in CO as $M_{\rm H}^{\rm DNM} = (62 \pm 7) {M_{\rm H_2}^{\rm CO}}^{\,0.51 \pm 0.02}$ across two decades in mass. }
{The phase transitions in these clouds show both common trends and environmental differences. These findings will help support the theoretical modelling of \hd formation and the precise tracing of \hd in the interstellar medium. }

\titlerunning{gas \& dust in nearby anticentre clouds II}

\maketitle

\section{Introduction}

Theoretical works on heating and cooling in the interstellar medium (ISM) predict the existence of two thermodynamically stable phases in the neutral atomic gas \citep{1969ApJ...155L.149F,1977ApJ...218..148M}: the Warm Neutral Medium (WNM) and the Cold Neutral Medium (CNM). The volume density and kinetic temperature in the WNM are in the range of 0.03--1.3 cm$^{-3}$, and 4100--8800 K, respectively, while in the CNM their ranges are 5-120 cm$^{-3}$, and 40--200 K, respectively \citep{2003ApJ...587..278W}. Given the densities of CNM and WNM, collisions between electrons, ions, and H atoms are able to thermalize the 21 cm transition in the CNM, but not in the WNM. So the spin temperature is expected to be close to the kinetic one in the CNM and less than the kinetic one in the WNM \citep{1985ApJ...290..578D,2001A&A...371..698L}. 
The 21 cm emission line of the atomic hydrogen traces the whole neutral atomic gas from WNM to CNM, but, without independent measurements of the 21 cm line in absorption toward distant radio sources to determine the spin temperature, one cannot retrieve the exact column density of hydrogen. Applying a uniform spin temperature, $T_S$, over a cloud complex or along sight lines provides an estimate of the proportion of CNM and WNM, and  only an average correction of the column densities in the dense CNM. The amplitude of column density correction with respect to the optically thin case increases with decreasing spin temperature and increasing \av. It reaches up to 35$\%$ at \av$\sim3$ \citep{2014ApJ...783...17L}.

The molecular phase is mainly composed of molecular hydrogen and is commonly traced by the $J\,{=}\,1\rightarrow0$ line emission of $^{12}$CO molecules at 115 GHz, hereafter simply referred to as CO lines. In their study of CO photo-dissociation and chemistry, \cite{1988ApJ...334..771V} predicted that CO emission may not trace all the molecular gas in translucent clouds. In the diffuse envelopes near the \hi-\hd transition, \hd molecules survive more efficiently than CO against dissociation by UV radiation \citep{2010ApJ...716.1191W}. Furthermore, the brightness of the $^{12}$CO line depends on the abundance and the level of collisional excitation of CO molecules, which are both low in the diffuse \hd envelopes. Large quantities of \hd can therefore be CO-dark below the sensitivity level of the current large-scale CO surveys (of order 1 \wcounit).

Chemical species precursors to CO formation, such as {C\,{\sc ii}}, CH, and OH, are used to trace the gas at the atomic-to-molecular transition.
\citet{2003ApJ...586.1111M} found that the CO line intensity in the MBM16 translucent cloud 
weakly correlates with the stellar reddening by dust, $E(B-V)$, contrary to the CH line intensity at 3335 MHz. Similarly, \cii line intensity at 158 microns and OH line intensity at 18 cm trace a larger fraction of molecular gas than CO line intensity in diffuse environments \citep{2010A&A...521L..18V,2010MNRAS.407.2645B}. However, \cii line emission also arises from the widely spread regions of atomic and ionized gas.

\cite{1990ApJ...352L..13B} have compiled a catalogue of clouds detected  by dust emission in the far infrared (FIR) at 100 $\mu$m with IRAS, but not seen in $^{12}$CO observations. These clouds at low extinction (<0.25 mag) were proposed to be made of diffuse molecular gas, here studied. \cite{2014A&A...564A..71R} have studied intermediate-velocity clouds (IVCs) with and without a FIR excess compared to the \hi column densities. They suggest that dim and bright IVCs in the FIR probe different stages of the transition from atomic to molecular phase.

Several studies have used the additional information provided by tracers of the total gas column density and have compared them to \hi and  $^{12}$CO line intensities in order to map the dark gas in the Galaxy at different scales.
It has been done using  \g-ray emission \citep{2005Sci...307.1292G}, thermal emission from large dust grains \citep{2011A&A...536A..19P}, and stellar reddening by dust \citep{2012A&A...543A.103P}.
In these studies, the \hi spin temperature is not fully constrained for each line of sight, so the additional gas may include a fraction of the CNM in addition to diffuse molecular gas. The suggestion that optically thick  \hi fully or largely accounts for the dark gas \citep{2015ApJ...798....6F} is challenged by \hi absorption, dust extinction, and \g-ray observations that we detail in the discussion part of the paper.
The additional gas is thus likely to include both optically thick \hi and CO-dark \hd at the transition between the atomic and molecular phases. In the absence of extensive emission tracers for this gas, we refer to this transition as the Dark Neutral Medium (DNM).

Simulations indicate that the CO line brightness rapidly saturates at large visual extinctions, near 4 magnitudes \citep{2011MNRAS.412.1686S}. Similarly, observations show that CO line absorption measurements at high optical depth reveal features not seen in emission  \citep{2012A&A...541A..58L}. 
In \cite{2017A&A...601A..78R} we have detected additional gas seen with both dust and \g-ray emissions toward dense regions where the $^{12}$CO line emission saturates. We refer to this CO-saturated, additional, \hd component as \cosat.

In this paper we use the DNM and \cosat maps derived jointly from \g-ray emission and dust optical depth at 353 GHz in addition to the \hi and CO emission data in order to investigate the \hi-\hd transition from diffuse to dense molecular clouds. We provide constraints on the range in \nhtot and \av required for \hd formation. We follow the evolution of the contribution of CO-dark \hd to the total molecular gas inside clouds of different types. We test the range of \hd column densities where $^{12}$CO line emission saturates.
In order to compare clouds of different types, we have selected a set of local clouds spanning between 140 and 420 pc in distance and with masses between 600 and 34400 solar masses. The sample includes the local anti-center clouds of Cetus, Taurus, Auriga, Perseus, California analysed in \cite{2017A&A...601A..78R} and the Chamaeleon cloud analysed in \cite{2015A&A...582A..31A}. These two analyses are based on the same gas tracers and same method.

The paper is organized as follows. Details on the method and on the cloud sample are given in the next section. The results and discussions are presented in Sec. 3. We focus on the transitions between the different gas phases in Sec. 3.1.  The evolution of the relative contributions of the CO-dark, CO-bright, and CO-saturated \hd to the molecular phase across a cloud is discussed in Sec. 3.2. Cloud-to-cloud variations of the average mass fractions are discussed in Sec. 3.3. The main conclusions and possible follow-up studies are listed in the last section.

 \begin{figure*}
  \centering                
  \includegraphics[scale=0.42]{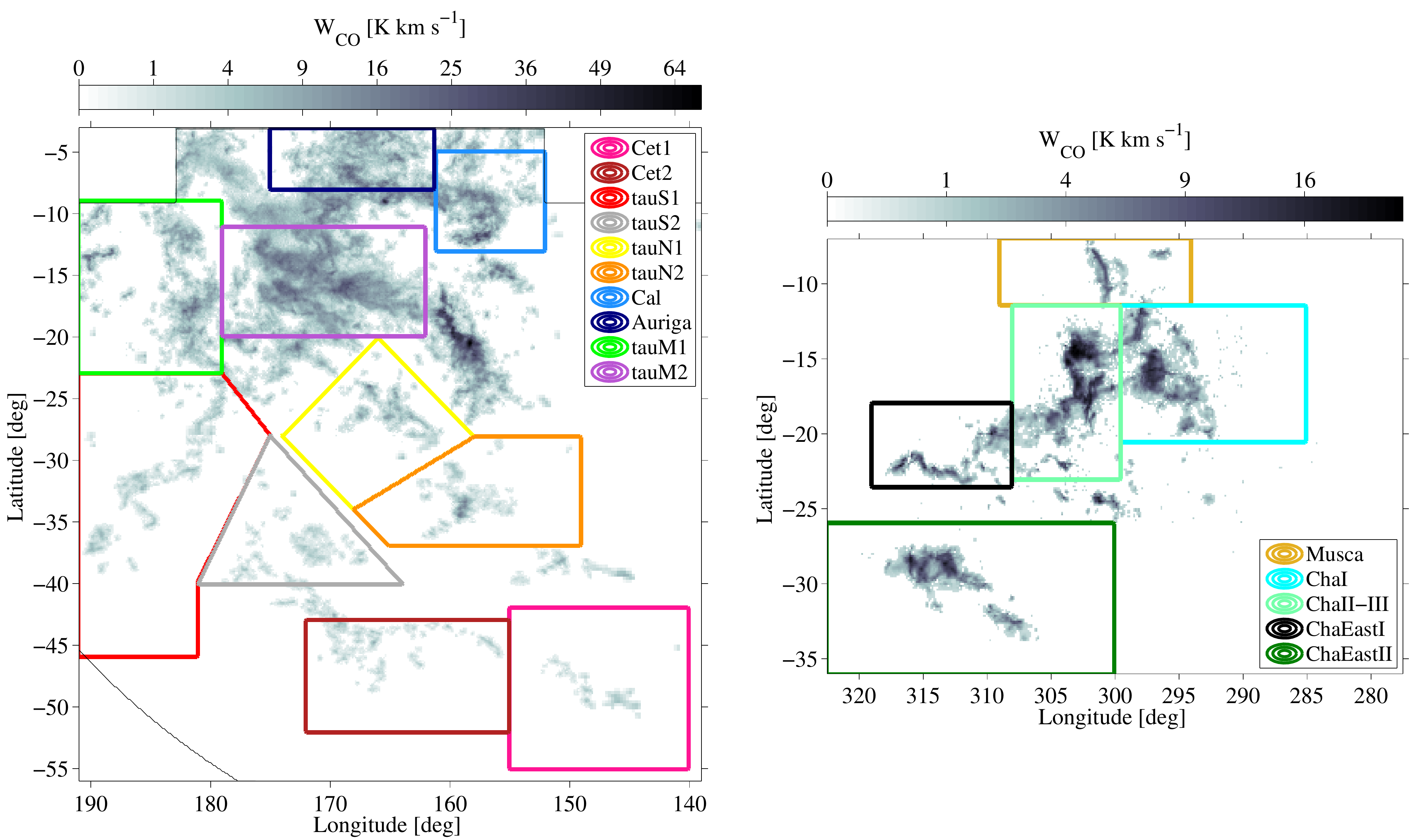}
  \caption{ Integrated CO intensity, \wco, overlaid with the boundaries of the selected substructures in the anticentre (left) and Chamaeleon (right) regions.}
   \label{fig:subRegions}
\end{figure*}

\section{Analyses and data}
The results presented here are based on the same dust and \g-ray analyses procedures as applied to two broad regions in the anticentre \citep{2017A&A...601A..78R} and in the Chamaeleon directions \citep{2015A&A...582A..31A} in the sky.
The analysis method and multi-wavelength data used to build the \g-ray and dust models are detailed in these papers, We summarize below their main features.

To trace the total gas we have used the dust optical depth, \taunu, inferred at 353 GHz from the spectral energy distribution of the thermal emission of the large grains, which has been recorded between 353 and 3000 GHz by \textit{Planck} and \textit{IRAS} \citep{2014A&A...571A..11P}.
We have also used six years of Pass 8 photon data from the \textit{Fermi} Large Area Telescope (LAT) between 0.4 and 100 GeV. Technical details on the photon selection, instrument response functions, energy bands, and the ancillary \g-ray data (local interstellar spectrum, point sources, inverse Compton and isotropic components) that we have used are given in \cite{2017A&A...601A..78R}.

In order to display the results as a function of the visual extinction \av in both regions, we have used the all-sky \ajj  map constructed by \citet{2016A&A...585A..38J} from the 2MASS extinction data, with the NICEST method at 12.0 arcmin resolution. \ajj values are converted into \av with a factor 3.55 according to the extinction law of \citet{1989ApJ...345..245C}.
Other \av datasets are available, but they do not cover both the anticentre and Chamaeleon regions using the same extraction method. We have not modelled \av as a total gas tracer as we have done with \g rays and \taunu because of significant discrepancies in the \av values obtained with different methods in the diffuse parts of the clouds (\av $\lesssim 1$ mag) that are the focus of this study \citep{2009MNRAS.395.1640R,2015ApJ...810...25G,2016A&A...585A..38J}.

\subsection{Summary of \g-ray and dust models} 
	
Galactic cosmic rays (CRs) interact with gas and low-energy radiation fields to produce \g rays. At the energies relevant for the LAT observations, the particle diffusion lengths in the interstellar medium exceed the cloud dimensions, and there is no spectral indication of variations in CR flux inside the clouds studied \citep{2015A&A...582A..31A,2017A&A...601A..78R}, so the interstellar \g radiation can be modelled, to first order, as a linear combination of gas column-densities in the various phases and the different regions seen along the lines of sight. The model also includes a contribution from the large-scale Galactic inverse-Compton (IC) emission, an isotropic intensity to account for the instrumental and extragalactic backgrounds, and point sources \citep[see Eq. 7 of][]{2017A&A...601A..78R}. 

The large dust grains responsible for the thermal emission seen in the far IR and at sub-mm wavelengths are supposed to be well mixed with the interstellar gas. For a uniform mass emission coefficient of the grains, the dust optical depth should approximatively scale with the total gas column densities, so we have also modelled the optical depth map as a linear combination of gas column-densities in the various phases and in the different cloud complexes \citep[see Eq. 6 of][]{2017A&A...601A..78R}.

The \g-ray intensity and the dust optical depth have been jointly modelled as a linear combination of \hi-bright, CO-bright, and ionized gas components (plus the non-gaseous ancillary components of the \g-ray model  mentioned above). The coefficients of the linear combination are used to scale the dust emission and \g-ray emission of gaseous origin into gas column density, and to probe independently the different gas phases and the different clouds separated in position-velocity. We have used 21-cm \hi and 2.6-mm CO emission lines to respectively trace the atomic and CO-bright molecular hydrogen, and we have used 70-GHz free-free emission to trace the ionised gas in the anticentre region. The contribution of ionized gas in the Chamaeleon region is negligible \citep{2015A&A...582A..31A}. The only way the spatial distributions of the \g-ray intensity and of the dust optical depth can correlate is via the underlying gas structure \citep{2005Sci...307.1292G,2015A&A...582A..31A}. So the joint information from dust emission and \g rays is used to reveal the gas not seen, or poorly traced, by {H\,{\sc i}}, free-free, and $^{12}$CO emissions, namely the opaque \hi and diffuse \hd present in the DNM, as well as the dense \hd to be added when the $^{12}$CO line emission saturates (\cosat). The two templates of additional gas used in the \g-ray model are inferred from the dust model, and conversely.
In the anticentre analysis we have separated the DNM and \cosat components in directions inside and outside of the 7 K km/s contour in \wco intensity (see Sec. \ref{sec:coldens} for details). Such a separation had not been done for the analysis of the Chamaleon region \citep{2015A&A...582A..31A}, so for the present study we have updated the analysis of the Chamaeleon region in order to separate in directions DNM and \cosat components using the same cut. 

The residual maps (data minus model) presented in \cite{2015A&A...582A..31A} and \cite{2017A&A...601A..78R} show that the best-fit models adequately describe both the \textit{Planck}-\textit{IRAS} observations in dust optical depth and the LAT observations in \g-ray intensity. Moreover jackknife tests have shown that the fitted parameters of the \g-ray and dust models apply statistically well to the whole region, both in the anticentre and Chamaeleon directions.

The residuals obtained between \g-ray data and the best fits are consistent with noise at all angular scales and they show that the linear model provides an excellent fit to the \g-ray data in the overall energy band, as well as in the separate energy bands. Significant positive residuals remain at small angular scales in the dust fit. They are likely due to the limitations in angular resolution and in sensitivity of the \g-ray DNM template used in the dust model. Small-scale clumps in the residual structure can also reflect localised variations in dust properties per gas nucleon that are not accounted for in the linear models. These effects are discussed in \cite{2017A&A...601A..78R}.

\input{DistMass_IG.txt}

\subsection{\hi and CO data and clouds selection}

\hi and CO emission in the anticentre and Chamaeleon regions have been mapped by different surveys. Their relative calibrations have been verified.
The Chamaeleon and anticentre analyses respectively use the CO data from the Nanten and Center for Astrophysics surveys \citep{2001PASJ...53.1071M,2001ApJ...547..792D,2004ASPC..317...66D}. After correction, the integrated CO intensities, \wco, recorded in the Chamaeleon region by the two surveys are close to a one-to-one correlation \citep{2015A&A...582A..31A}.
The \hi survey GASS II \citep{2010A&A...521A..17K} covers the Chamaeleon region. The GALFA and EBHIS surveys \citep{2011ApJS..194...20P,2016A&A...585A..41W}  encompass the anticentre region. We have verified the good correlation between EHBIS and GALFA column densities in the anticentre region \citep{2017A&A...601A..78R}. According to \cite{2015A&A...578A..78K} and \cite{2016A&A...585A..41W}, the brightness temperature of GASS II is on average 4\% higher than that of EBHIS. This discrepancy does not affect the column density profiles we present through single clouds (each covered by a single survey), but it induces a small shift between the gas fractions measured in the Chamaeleon region relative to the anticentre ones. Given the uncertainties, of order 5 to 10\%, in the \g-ray and dust models, a 4$\%$ shift does not significantly bias our results.
We can therefore compare the \hi column densities and CO integrated intensities measured in both regions and merge the two samples of clouds to study cloud-to-cloud variations.

\begin{figure*}[p!]
  \centering 
  \includegraphics[width=\hsize]{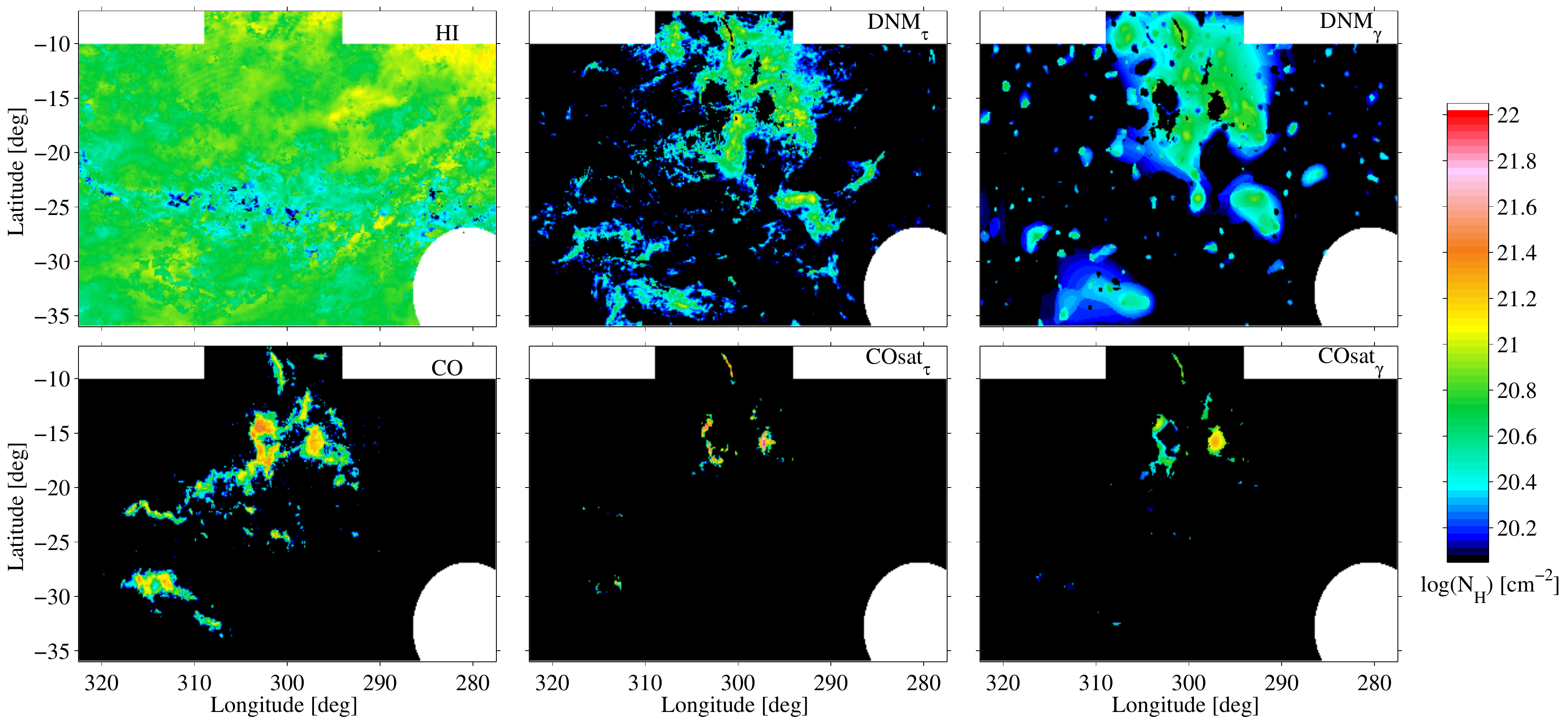}
\caption{  Hydrogen column density maps of the HI, DNM (derived from the \taunu and \g-ray fits), CO-bright, and \cosat (derived from the \taunu and \g-ray fits) components in the Chamaeleon region. The \xco and \opa scale factors used in their derivation come from the \g-ray analysis for optically thin \hi.\vspace{0.8cm}}
   \label{fig:NHcha}
\end{figure*}

\begin{figure*}[p!]
  \centering 
  \includegraphics[width=\hsize]{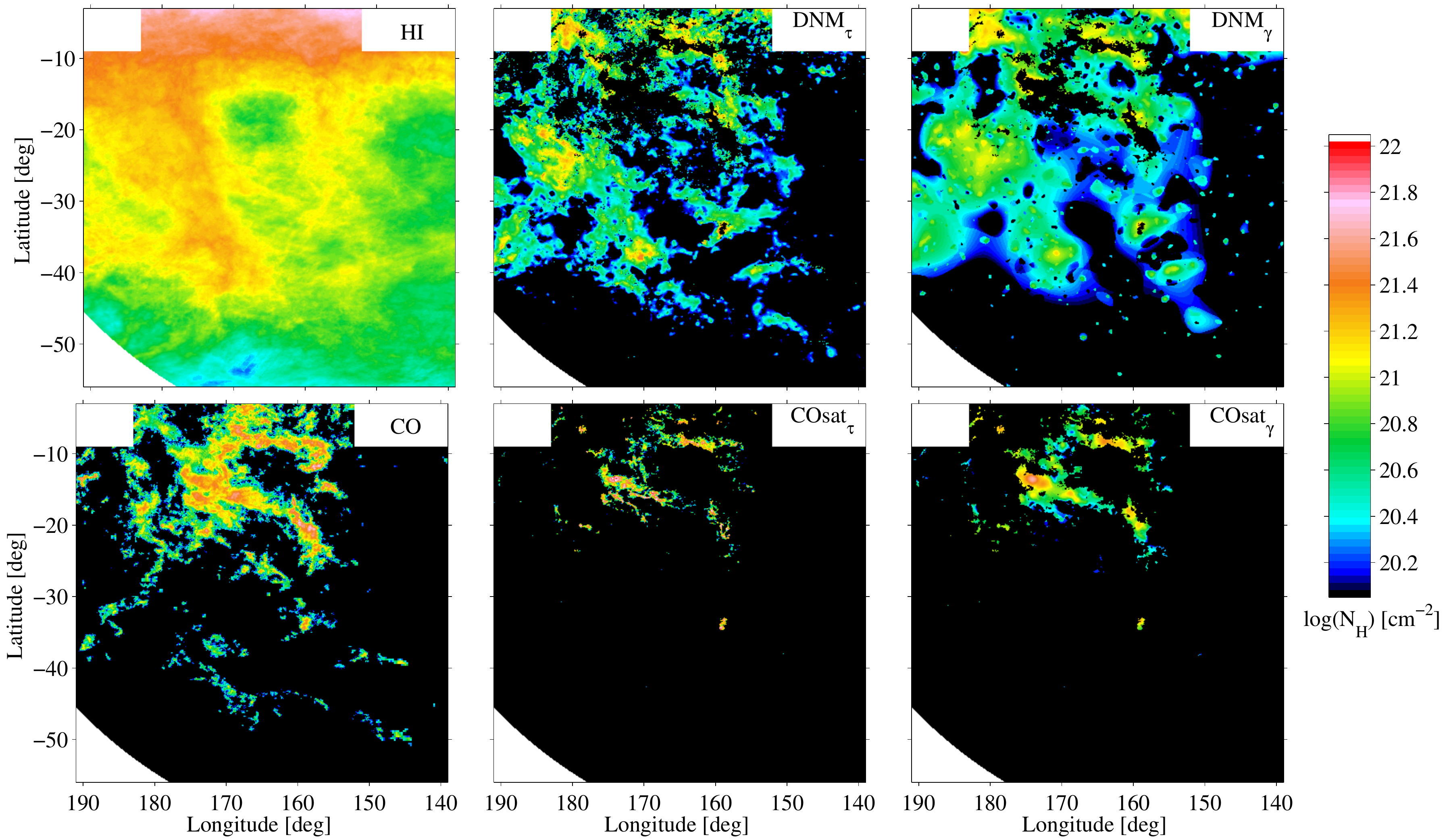}
\caption{  Hydrogen column density maps of the HI, DNM (derived from the \taunu and \g-ray fits), CO-bright, and \cosat (derived from the \taunu and \g-ray fits) in the anticentre region. The \xco and \opa scale factors used in their derivation come from the \g-ray analysis with an \hi spin temperature of 400K.\vspace{0.8cm}}
   \label{fig:NHtau}
\end{figure*}

 \begin{figure*}
  \centering   
  \hspace{0.5cm}               
  \includegraphics[scale=0.33]{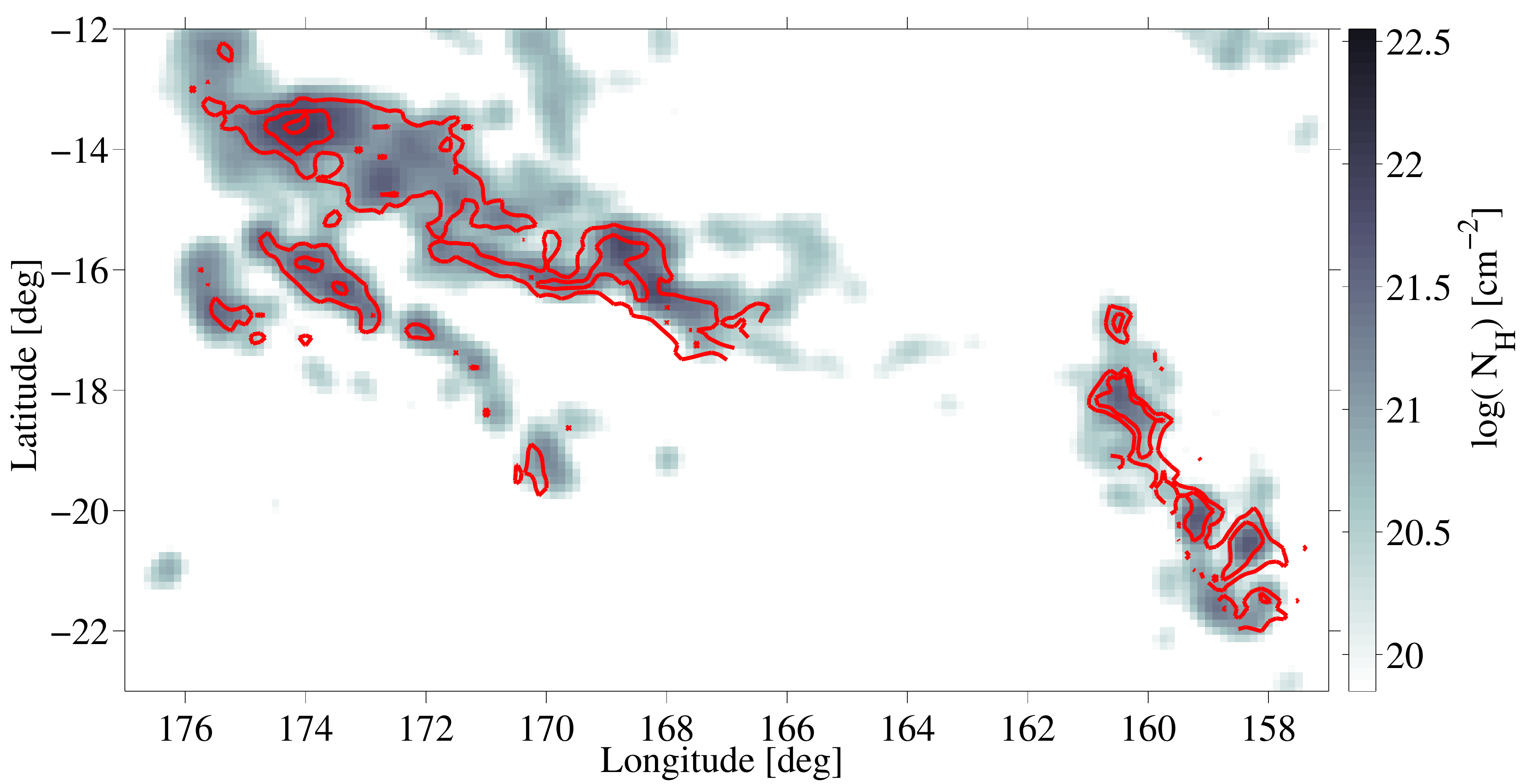}
  \hspace{0.5cm}  
  \caption{ Hydrogen column density map of the \cosat component from the \taunu analysis overlaid with contours of $^{13}$CO line intensities at 2 and 4 K km s$^{-1}$ in the Taurus and Perseus clouds \citep{2008ApJS..177..341N,2006AJ....131.2921R}. \vspace{0.8cm} }
   \label{fig:DNM13co}
\end{figure*}

We have used \hi and CO lines to kinematically separate cloud complexes along the lines of sight in both regions and to separate the nearby clouds from the Galactic backgrounds. Details of the pseudo-Voigt line decompositions and selection of longitude, latitude, and velocity boundaries of the complexes are given in \cite{2015A&A...582A..31A} and \cite{2017A&A...601A..78R}. The analyses resulted in the separation of six different complexes in the anticentre region  \citep[Cetus, South Taurus, North Taurus, Main Taurus Perseus, and California, see Fig. 1 of ][]{2017A&A...601A..78R} and two complexes in the Chamaeleon region \citep[Chamaeleon, IVA, see Fig. 2 of ][]{2015A&A...582A..31A}. The complexes separated in position-velocity form coherent entities in \hi and $^{12}$CO to which we can associate specific distances compiled from the literature \citep{2015A&A...582A..31A,2017A&A...601A..78R}.

We have selected 15 clouds or sub-regions within the broader complexes in order to study phase transitions in different entities. The contours and names of the sub-regions are given in Fig. \ref{fig:subRegions}. When referring to those clouds, we will propagate the same names and colour scheme throughout the paper. The set of sub-regions in the Chamaeleon region follow that studied in \cite{2015A&A...582A..31A}. The boundaries of the sub-regions have been chosen to avoid directions where the different complexes overlap along the lines of sight, so that we can study the relative contributions of the different gas phases to the total column density. This is particularly important for the DNM contributions which cannot be kinematically separated along the lines of sight. Inside a given cloud area, we associate all the DNM gas to the cloud. The morphological coherence between the  \hi, DNM,  and CO structures within the area strengthens the DNM association. The areas around the CO clouds are wide enough to encompass the extent of their DNM phase. One cannot rule out the presence along the line-of-sight of an isolated translucent (DNM) cloud without CO counterpart in addition to the DNM envelopes surrounding the visible CO clouds. However recent simulations \citep{2016A&A...587A..76V} have shown that the different gas phases are well intermixed, so we expect most of the DNM gas to be spatially associated with the \hi and CO phases present in the area.

We do not attempt to leave the parameters of the dust and \g-ray models free in the 15 smaller clouds because of the current \g-ray statistics, because of the separation in velocity of the different complexes, and because of the broad extension of the local \hi structures. We obtain excellent \g-ray residuals with only six local complexes, so there is no obvious need for further subdivision. One should also note that the 15 sub-regions do not cover the whole analysis regions because we have excluded directions where clouds overlap in velocity/distance. It is not possible to model the 15 clouds without also modelling the rest of the regions.
 
The 15 sub-regions are presented in Table \ref{tab:CLmass}, together with their masses inferred from the \hi and CO data, their star-formation rates (SFR) and their \xco factors. All quoted masses are directly derived from the column density maps, taking into account the Helium contribution. The \hi masses have been calculated for optically thin conditions in the Chamaeleon region and for a spin temperature of 400 K in the anticentre region. Those conditions best match the interstellar \g-ray maps. We have applied a different CO-to-\hd conversion factor, \xco $=$\nhd/\wco, for each of their parent complexes (as we do not leave the model parameters free in the 15 clouds). The SFR have been derived from the mass of dense molecular gas seen in CO emission at \av >8 mag according to the empirical relation expected from the ``microphysics'' of prestellar core formation within filaments \citep{2014prpl.conf...27A,2015A&A...584A..91K,2017arXiv170500213S}:\\ \mbox{SFR=$(4.5\pm2.5) \times 10^{-8}$ M$_{\odot}$ yr$^{-1}$ $\times$ (M$^{A_{\rm V}>8}_{\rm{H_2}} $/M$_{\odot}$)}. This relation is in good agreement with measurements in Local clouds  \citep{2010ApJ...724..687L,2017arXiv170500213S}.  Table  \ref{tab:CLmass} shows that the dataset includes clouds ranging from low column density translucent clouds to star-forming molecular clouds belonging to the solar neighbourhood, with distances ranging between 140 and 420 pc and with a large span of masses between 200 and 18300 solar masses in the CO-bright phase.


\subsection{Estimation of the gas column density}\label{sec:coldens}

Several studies have shown that the dust opacity, \opa, rises as the gas, whether atomic or molecular, becomes denser. This opacity increase is likely due to a change in emission properties of the dust related to a chemical or structural evolution of the grains \citep{2012ApJ...751...28M,2015A&A...579A..15K}. In the dense molecular filaments of the Taurus and Chamaeleon clouds the opacity can rise by a factor of 2 to 4 above the value found in the diffuse ISM \citep{2003A&A...398..551S,2011A&A...536A..25P,2013A&A...559A.133Y,2015A&A...582A..31A,2017A&A...601A..78R}. Such a rise severely limits the use of the dust thermal emission to trace the total gas in dense media. The limit of the linear regime is estimated to be \nhtot$<2\times10^{21}$cm$^{-2}$ in the Chamaeleon clouds and  \nhtot$<3\times10^{21}$cm$^{-2}$ in the anticentre clouds.

In all the nearby clouds studied so far and across all the gas phases we find no evidence of spectral evolution of the \g-ray emissivity of the gas relative to the local average. The latter is referred to as the Local Interstellar Spectrum (LIS). This consistency is verified in the denser molecular parts (\cosat). So at the multi-GeV to TeV particle energies relevant for the LAT observations, the same CR flux pervades the clouds, from the diffuse atomic outskirts to the dense parts seen in $^{12}$CO line emission. Therefore the same CR flux pervades through the bulk of a cloud volume and mass. The \g-ray emission efficiently traces the total column density of gas, contrary to the dust optical depth which is affected by the evolution of the grain emission properties. So we will preferentially estimate \nhtot using \g rays. 

The total column density is  the sum of the {H\,{\sc i}}, DNM, CO, and \cosat contributions scaled according to the best-fit parameter of the \g-ray model (details for each components are given in the following paragraphs).  The total molecular column density is the sum of the molecular DNM, CO, and \cosat column densities derived from the \g-ray fit: \mbox{$N_{\rm{H_2}}^{\rm{tot}}=N_{\rm{H_2}}^{\rm{DNM}}+N_{\rm{H_2}}^{\rm{CO}}+N_{\rm{H_2}}^{\rm{COsat}}$}. As the DNM gas can be composed of both opaque \hi and diffuse \hd, we have assumed a composition varing between 50 and 100\% of molecular hydrogen (see discussion at the beginning of Sec. \ref{sec:dkCO}). By default the results are given for a half molecular DNM composition, except if specified otherwise.

For the calculation of the \hi column densities, \nhi, we did not attempt to set a different \hi spin temperature for each complex, but we have used a uniform average spin temperature. In the following the calculations are performed at the best-fit spin temperature of 400 K in the anticentre clouds, and for an optically thin emission in the Chamaeleon clouds \citep{2015A&A...582A..31A,2017A&A...601A..78R}. 

The column density map of DNM component map derived from the dust distribution and \taunu analysis has been scaled in mass with the \g-ray emission under the assumption of a uniform CR flux permeating the \hi and DNM phases. This assumption is substantiated by the uniformity of the CR spectrum found in the {H\,{\sc i}}, DNM, and CO phases. By construction, the DNM maps exclude spatially the denser CO regions (\wco > 7 K km s$^{-1}$) that are attributed to the \cosat component. This separation we have adopted between the DNM and \cosat components is only a directional boundary. The 7 K km s$^{-1}$ cut in \wco intensity is chosen deeply enough into CO clouds to ensure that the CO-bright \hd column density dominates the \nh in the other phases. This choice is specific to the present set of clouds to ensure a uniform cut in all 6 local cloud complexes. Because of the complex 3D structure of the clouds, the DNM can be detected along lines of sight intercepting CO emission in excess of 1 K km s$^{-1}$, but the fraction of DNM gas decreases from the cloud outskirts to its dense, CO-intense parts. The choice of 7 K km s$^{-1}$ ensures that the additional gas detected by the dust and \g rays toward the dense molecular regions is predominantly, in column density, in the \cosat phase. We show in Sec. \ref{sec:res} that the DNM gas contributions toward these directions do not exceed 10\% of the \cosat column density.

The quantity of gas in the molecular phase is usually estimated via the $^{12}$CO line emission and scaled into hydrogen column density via the \nhd/\wco conversion factor, \xco. Thanks to the good penetration of CRs up to the CO-bright phase, the \hi and CO related parameters of our \g-ray model allow the derivation of mean \xco values for each cloud complex independently. In the previous paper \citep{2017A&A...601A..78R}, we have shown that these average \xco ratios significantly decrease from the diffuse clouds, subject to heavier photo-dissociation, to the more compact, dense, and well-shielded CO clouds. According to theory \citep{2006MNRAS.371.1865B,2011MNRAS.412..337G}, the differences in mean \xco values reflect intrinsic differences in CO abundance and excitation conditions, therefore in \xco spatial gradients across the various clouds. This is why we have used the individual \xco values of Table \ref{tab:CLmass} in our models. We have not attempted to model spatial gradients in \xco across the clouds since there is no consensus on the theoretical prescription for how \xco should scale with \wco \citep{2006MNRAS.371.1865B,2010AeA...518A..45L,2011MNRAS.412..337G,2011MNRAS.412.1686S,2012AeA...541A..58L,2016MNRAS.455.3763B}.

\begin{figure*}[p!]
  \centering                
  \includegraphics[scale=0.467]{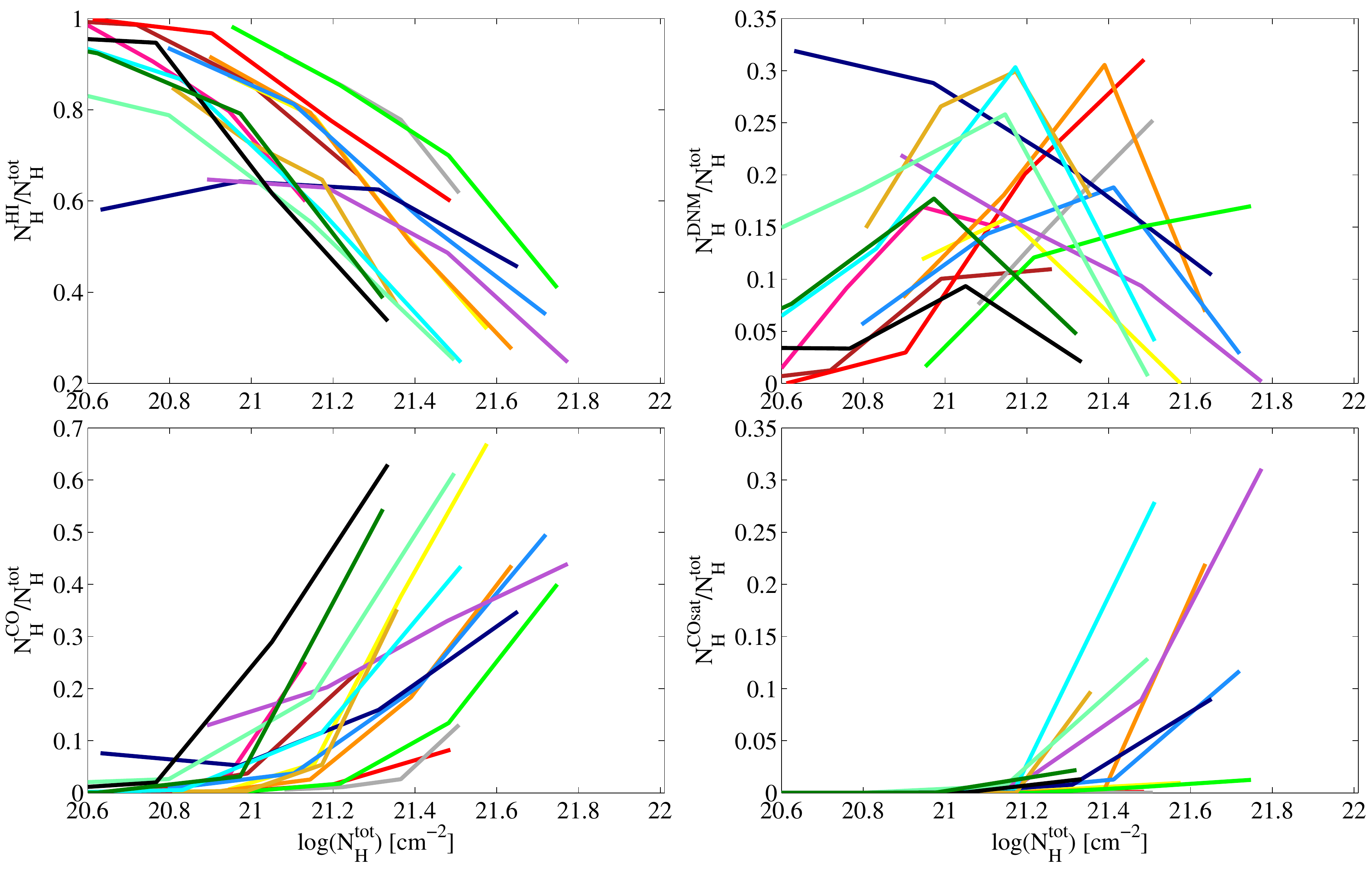}
  \caption{ Evolution with \nhtot of the fractions of the total hydrogen column density in the \hi (top left), DNM (top right), CO (bottom left), and \cosat (bottom right) components. The number of points is the same for all the curves. The points are equally spaced between the minimum and maximum of each profile. The colors refer to the same clouds as in Fig. \ref{fig:subRegions}. }
   \label{fig:9phases}
\end{figure*}

\begin{figure*}[p!]
  \centering                
  \includegraphics[scale=0.467]{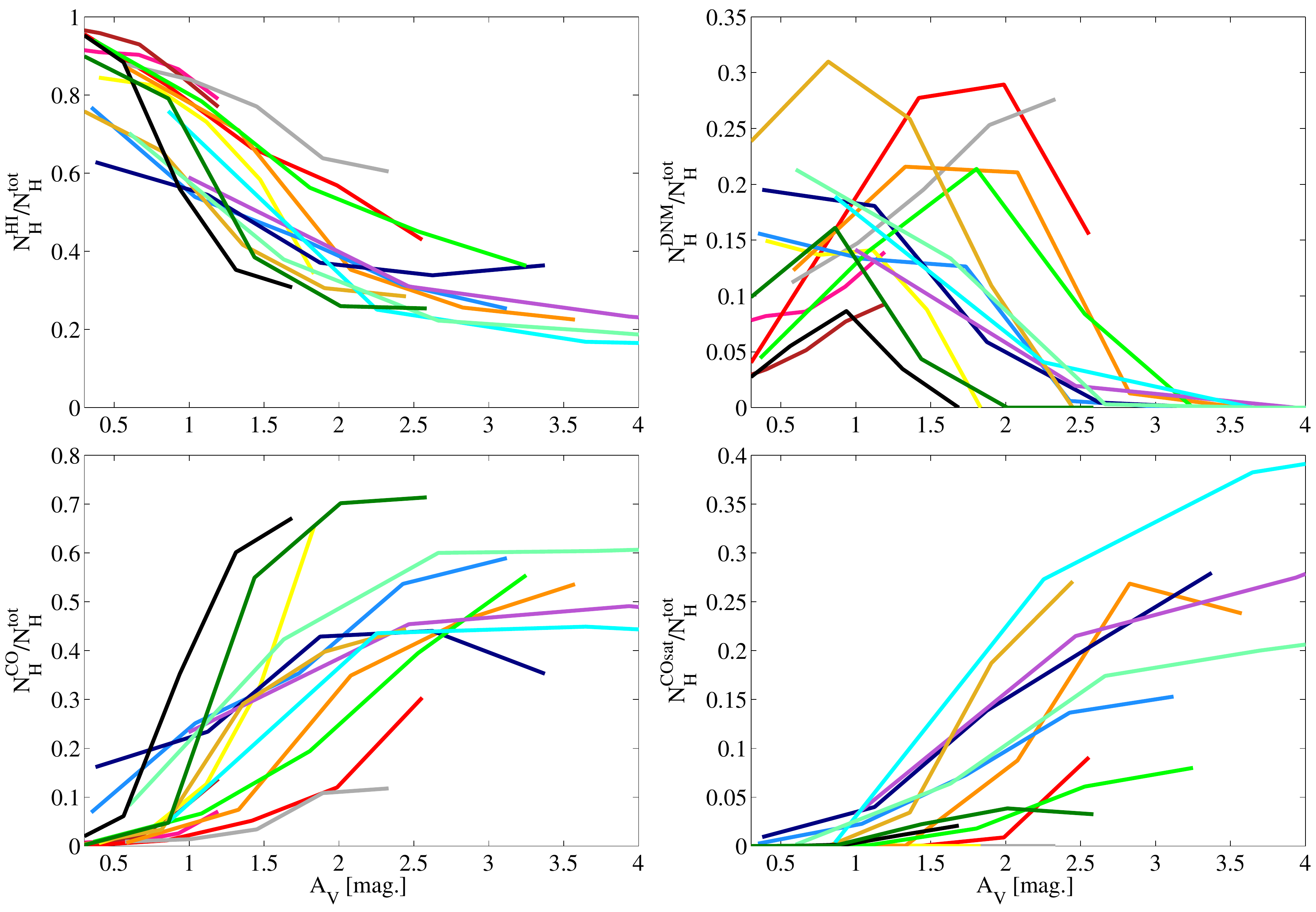}
  \caption{ Evolution with \av of the fractions of the total hydrogen column density in the \hi (top left), DNM (top right), CO (bottom left), and \cosat (bottom right) components. The colors refer to the same clouds as in Fig. \ref{fig:subRegions}.}
   \label{fig:9phasesAV}
\end{figure*}

\section{Results and discussions}\label{sec:res}


The results of the \g-ray and dust models obtained from the updated analysis of the Chameleon region compare very well with the published ones.
The \xco factor of (0.65$\pm$ 0.02) \xcounit derived from the \g-ray fit has not changed.
The new dust opacities found in the {H\,{\sc i}}, DNM, and CO phases are respectively \opavghi=(15.0$\pm$ 0.2) \opaunit, \opavgdnm =(17.9$\pm$ 0.5)  \opaunit, and \opavgco=(32.9$\pm$ 1.0)  \opaunit. Compared to values from the previously cited analyses the $-8$\%, $-4$\%, and $+3$\% changes are not significant. 

The hydrogen column-density maps obtained in the {H\,{\sc i}}, DNM, CO, and \cosat phases seen in the anticentre and Chamaeleon regions are shown in Fig. \ref{fig:NHcha} and \ref{fig:NHtau}, respectively. The CO map is based on the \wco data and on the \xco ratios derived in \g rays for each individual cloud complex. In both regions, the column densities span from 10$^{20}$ cm$^{-2}$  in the atomic gas to a few times 10$^{22}$ cm$^{-2}$ in the molecular cores. We find comparable ranges of column densities in each phase of neutral gas. The means of the column density distributions in the anticentre clouds are larger than in the smaller Chamaeleon clouds, typically by a factor of two in the \hi and DNM phases, and by a factor of three in the CO and \cosat phases. 

Figures \ref{fig:NHcha} and \ref{fig:NHtau} illustrate that both analyses exhibit significant quantities of gas not linearly traced by \nhi column densities and \wco intensities, but jointly revealed by the dust mixed with gas and by the CRs spreading through it. 
The separation of the additional gas into DNM and \cosat components according to the \wco intensity in each direction (for \wco $>$ 7 K km s$^{-1}$) highlights the structural differences between the spatially extended, diffuse DNM gas gathering at the \hi and CO interface, and the more compact, denser filaments and clumps of molecular gas in regions of large \wco intensities. 
Figure \ref{fig:DNM13co} shows that the additional gas detected in the \cosat component is likely due to large CO opacities because of the good spatial correspondence between the distributions of the \cosat hydrogen column densities and of the intensities of the optically thinner $^{13}$CO $J\,{=}\,1\rightarrow0$ lines observed in the main Taurus and Perseus clouds \citep{2008ApJS..177..341N,2006AJ....131.2921R}. 
Further exploiting the quantitative relation between the observed $^{13}$CO intensities, the saturating $^{12}$CO intensities, and the \cosat column densities is beyond the scope of this paper. It requires a careful modelling of the $^{12}$CO and $^{13}$CO radiative transfer and of the non-linear evolution of the dust emissivities that are used with \g rays to trace the additional gas (see the discussions of the evolution of \opa opacities at large \nhtot in \cite{2017A&A...601A..78R}). 
We note that clouds with equivalent \wco intensities may have different quantities of additional \hd in the \cosat component. For example, we observe comparable \wco intensities in the filaments of Musca and ChaEastI, and in the clumps of ChaI and ChaEastII, but we find hardly any \cosat gas in ChaEastI and ChaEastII whereas Musca and ChaI apparently require additional column densities as large as $10^{22}$~cm\msq.

The use of a $^{13}$CO template covering the whole region would add a morphological constraint on the fit to trace the dense molecular gas not seen in $^{12}$CO emission. This would greatly help to separate \cosat from DNM along the lines of sight. Moreover it would particularly improve the fit of the dust model in the dense \hd regions where the flaring of the \opa ratios or, alternatively, the loss of precise E(B-V) measurements can significantly bias the estimation of the gas column density from the dust tracers.
In the absence of an independent template for \cosat, such as $^{13}$CO emission, our analysis cannot separate the additional \cosat gas from the foreground and background DNM present along the lines of sight toward the dense regions of large \wco intensities. So the column densities of our \cosat map include a small contribution from the DNM gas. The interpolation of the DNM \hd column densities across the \cosat regions indicates that the \cosat column density could be lowered by about 5--10$\%$ after subtraction of the DNM contributions.

DNM structures spatially extend between those of \hi and CO clouds. The comparison between different clouds in the maps show environmental or evolutionary differences in the DNM content of clouds. This is illustrated by differences among the high-latitude translucent molecular clouds. MBM18 ($l=189$\fdg1 $b=-36\degree$), MBM16 ($l=170$\fdg6 $b=-37$\fdg3), MBM12 ($l=159$\fdg4 $b=-34$\fdg3), and the chain of CO clouds along Cetus (at $b<-40\degree$) exhibit a rich DNM component contrary to MBM8 ($l=15$1\fdg7 $b=-38$\fdg7) and MBM6 ($l=145$\fdg1 $b=-37$\fdg3).
The proportion of optically thick \hi and CO-dark \hd in the DNM composition may also vary from cloud to cloud.
An interesting structure is the DNM complex located at $170\degree<l<190\degree$ and $-40\degree<b<-20\degree$. This is one of the largest DNM structures identifiable in the early all-sky DNM maps \citep{2005Sci...307.1292G,2011A&A...536A..19P,2012A&A...543A.103P}. It extends spatially well beyond the edges of CO emission with column densities $10^{20}$<\nhdnm<$10^{21}$ cm$^{-2}$. Its morphology is comparable to that of the \hi gas in South Taurus. Its structure contrasts with the DNM found close to the CO edges of the MBM clouds and the California cloud where the column densities can reach larger values near 2$\times10^{21}$ cm$^{-2}$. We note that the fraction of the shell in California cloud missing in CO at $l=159$\fdg5 and $b=-10\degree$ is filled by DNM gas. 

\begin{figure}
  \centering                
  \includegraphics[width=\hsize]{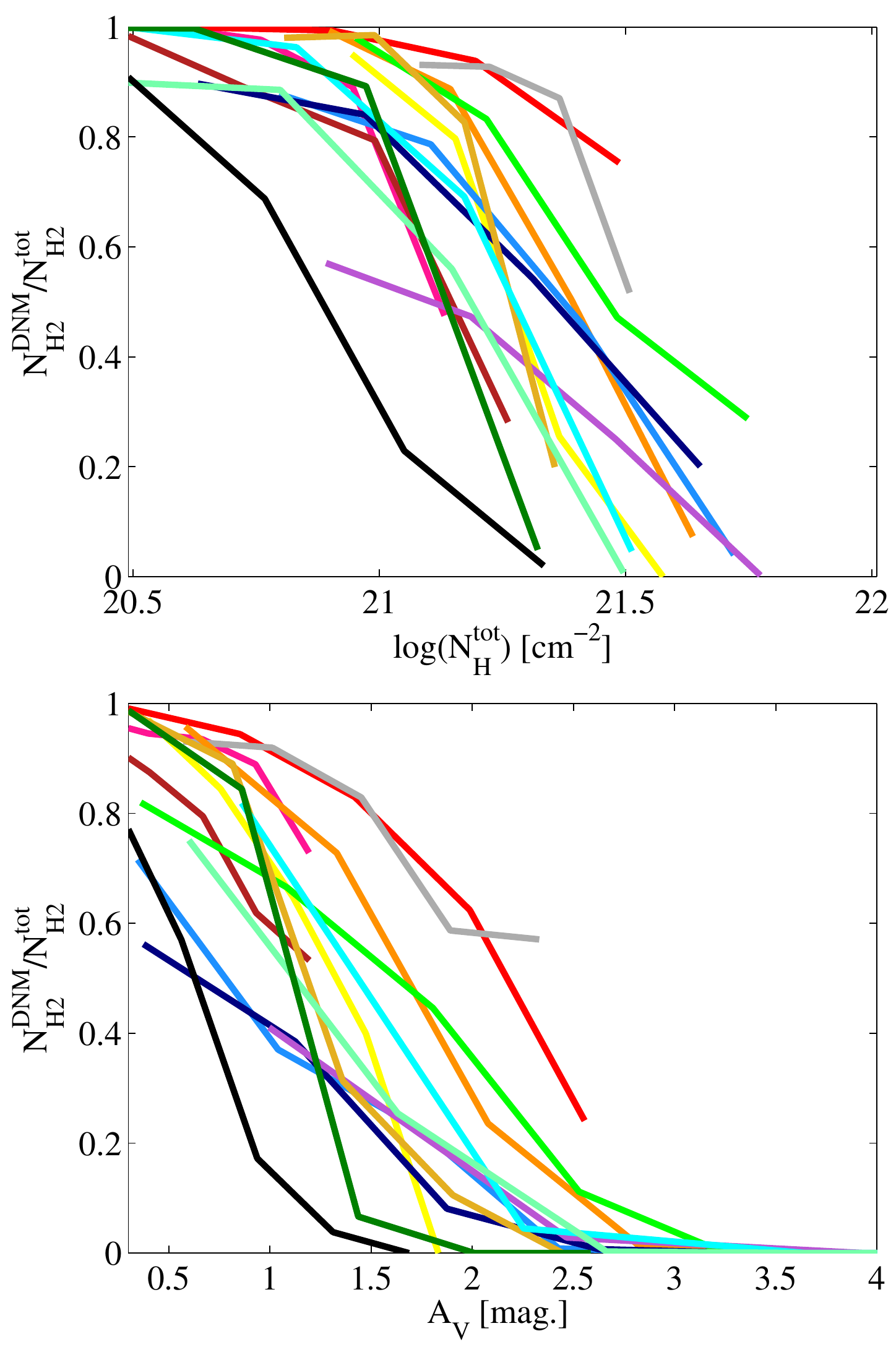}
  \caption{ Evolution of the fraction \fcodk of molecular DNM in the total \hd column-density as a function of the total gas column density \nhtot (top) and visual extinction \av (bottom). The DNM gas is assumed to be $50\%$ molecular. The colors refer to the same clouds as in Fig. \ref{fig:subRegions}.}
   \label{fig:9nhdark}
\end{figure}

\subsection{Phase transitions}\label{sec:transi}

In order to study the transition between the different neutral gas phases we have derived the ratios of the hydrogen column densities in the {H\,{\sc i}}, DNM, CO-bright, and \cosat components over the total column density.
The ratios do not depend on cloud distances.
We have plotted the ratios in the four neutral gas phases as a function of the total column density \nhtot and of the visual extinction \av in Figs. \ref{fig:9phases} and \ref{fig:9phasesAV}, respectively. Because of systematic uncertainties in the current measurements of low extinctions, we do not attempt to study the phase transitions below 0.3 magnitude in \av. 

We find that in all fifteen clouds the \hi fraction in the total column density decreases for column densities greater than (0.6-1)$\times$10$^{21}$ cm$^{-2}$, in agreement with the theoretical limit of 10$^{21}$ cm$^{-2}$ necessary to shield H$_{2}$ against UV dissociation for clouds of solar metallicity according to the models of \citet{2009ApJ...693..216K} and \citet{2014ApJ...783...17L}.
This limit is also confirmed by FUSE observations of H$_{2}$ \citep{2006ApJ...636..891G}, by OH observations \citep{2010MNRAS.407.2645B}, and by dust observations in Perseus  \citep{2012ApJ...748...75L}, which measure \nhi column densities leveling off at (0.3-0.5), (0.4-0.5), (0.8-1.4) $\times$ 10$^{21}$ cm$^{-2}$, respectively.  We note that, even though the \hi mass of the individual clouds spans a tenfold range, the \hi fractional decrease occurs over a small range in \nhtot, only a factor of four wide from cloud to cloud. We also note that the slope of the decline is approximately the same in the different clouds. The column density fractions in the \hi often remain large, about 30\%, toward the CO-bright phase because of the turbulent 3D structure of the clouds \citep{2016A&A...587A..76V} and because of the atomic envelopes surrounding the molecular clouds.

The CO-dark and CO-bright regimes in clouds are expected to be offset in depth because \hd is better shelf-shielded against UV dissociation than CO at low extinction. The combination of dust screening and gas shielding is required at large extinction to protect CO molecules \citep{2009A&A...503..323V}. The profiles of Fig. \ref{fig:9phasesAV} show transitions to CO-bright emission that tend to be deeper into the clouds than for the \hi-\hd transition. At \nh<10$^{21}$ cm$^{-2}$ (\av$<$1 mag), the decrease in \hi column-density fraction is larger than the increase in CO fraction. In this range the DNM fractions exceed  the CO ones.

The DNM fraction in the total column density often peaks in the (1-3)$\times$ 10$^{21}$ cm$^{-2}$ range. Compared to the higher degree of organisation of the \hi and CO-bright transitions, the DNM fractions show a large diversity of profiles that may be due to the large spatial overlap between the HI-bright, DNM, and CO-bright phases when integrated along the lines of sight.

Note that the DNM column densities are not very sensitive to the sensitivity threshold in CO observations for two reasons. Firstly, optically-thick \hi contributes with diffuse CO-dark \hd to the DNM column density. The \hi proportion is unknown, but likely not zero. Secondly, a large fraction of the DNM map would have easily been detected in $^{12}$CO emission if the DNM gas were mostly \hd with a CO abundance and level of excitation equivalent to the conditions prevailing in the CO-bright parts. In order to check this assertion, we have applied an \xco ratio of \xcounit, which characterizes the diffuse CO clouds in our sample, to translate the DNM column densities to equivalent \wco intensities. We have then calculated the fraction of the total DNM solid angle where the equivalent \wco intensity exceeds 1 \wcounit level that would have been easily detected. We find a fraction of 58\% in solid angle, which gathers 74\% of the DNM mass. In those directions the lack of CO emission is not due to the sensitivity limit of the CO survey, but to the genuine under-abundance and/or under-excitation of CO molecules.

The CO sensitivity threshold of the present observations are respectively 0.4 and 1.5 K km s$^{-1}$ for the anticentre and Chamaeleon regions \citep{2001ApJ...547..792D,2015A&A...582A..31A}. At that sensitivity level, the onset of CO intensities occurs for total gas column densities ranging between 0.6 and 2.5 $\times$ 10$^{21}$ cm$^{-2}$, in close correspondence with the transitional drop in \hi. We note again a small, factor-of-four, variation in the CO-bright transition from cloud to cloud.

The CO-bright transition occurs at total \nhd column densities of (1.5-4)$\times$ 10$^{20}$ cm$^{-2}$.
Simulations of photo-dissociation regions (PDR) by \cite{2012A&A...544A..22L} show that the column density of CO molecules rises steeply after the molecular transition, at hydrogen volume densities of about 100 cm$^{-3}$ and at \nhd=(3-6)$\times$ 10$^{20}$ cm$^{-2}$ depending on the density profile in the PDR region. A uniform density profile shifts the transition to larger \nhd.  
The value of 3$\times$ 10$^{20}$ cm$^{-2}$ corresponding to the non-uniform case agrees reasonably well with our results, as well as with CO and \hd observations by the \textit{HST} and \textit{FUSE}, which indicate a transition at \nhd=2.5 $\times$ 10$^{20}$ cm$^{-2}$ \citep{2008ApJ...687.1075S}.

The saturation of the CO emission lines corresponds to a fractional increase of the column densities in the \cosat component, but they contribute only 10\% to 30\% more gas. The saturation occurs in the main molecular clouds beyond (1.5-2.5)$\times$ 10$^{21}$ cm$^{-2}$ in \nhtot.




\subsection{Evolution of the CO-dark \hd fraction in column density within the clouds}\label{sec:dkCO}

This section focuses on the transitions inside the molecular phase from the CO-dark to CO-bright and then CO-saturated regimes in each of the clouds in our sample. We distinguish the diffuse CO-dark \hd present in the DNM from the dense \hd implied by the \cosat component to the bright CO cores in order to isolate the fraction of dark molecular gas lying at the diffuse atomic-molecular interface. 

We first discuss the transition from CO-dark to CO-bright \hd by measuring the change in the fractions of molecular DNM in the total \hd column density, \nhd.  We define the CO-dark \hd fraction in column density as:
\begin{align}
f_{\rm{COdk \, H_2}} = \frac{N_{\rm{H_2}}^{\rm{DNM}}}{N_{\rm{H_2}}^{\rm{tot}}}=\frac{N_{\rm{H_2}}^{\rm{DNM}}}{N_{\rm{H_2}}^{\rm{DNM}}+N_{\rm{H_2}}^{\rm{CO}}+N_{\rm{H_2}}^{\rm{COsat}}}
\end{align}
Instead of the generic ``dark gas fraction'' ($f_{DG}$) term found in the literature for column density or mass ratios and which often implicitly assumes that the whole DNM is molecular, we use the more explicit designation of ``CO-dark \hd fraction'' to restrict the fraction to the molecular part of the DNM (excluding the opaque \hi) and to exclude issues with saturation of the \wco intensities (\cosat components). 

As the DNM lies at the interface between the atomic and molecular phases, it is expected to be composed of optically thick \hi and diffuse \hd (not seen in CO at the level of sensitivity used here), both mixed with a proportion varying as the gas becomes denser and according to the ambient conditions \citep[e.g. \hd volume density and kinetic temperature, intensity  of the interstellar radiation field, ISRF, dust content, flux of ionizing low-energy CRs, ][]{2009A&A...503..323V}.
The suggestion that optically thick \hi fully or largely accounts for the additional gas in the DNM \citep{2015ApJ...798....6F} is challenged by several observations. 
\citet{2015ApJ...798....6F} have estimated a spin temperature of T$_S$=20–40 K and an optical depth $\tau_{max}$>0.5 for 85\% of their sky coverage (|$b$| > 15\degree).  \cite{2014ApJ...793..132S} have measured \hi absorption against 26 radio continuum sources toward the Perseus cloud and have found that 54\% of directions have $\tau_{max}$>0.5 and only 15\% of lines of sight have a spin temperature lower than 40 K. Their observation suggests that \hi absorption traces mostly the central cloud regions where CO is bright, and only to a smaller degree the CO-dark envelopes, where DNM gas is more abundant (see Figs. \ref{fig:NHcha} and \ref{fig:NHtau}).
Corrections of the \hi column densities are too small to fully explain the changes observed in the \nhi/$E(B-V)$ ratios at $E(B-V)>0.1$ so larger column density must be attributed to the onset of \hd formation \citep{2014ApJ...783...17L}.
Analysis of the dust thermal emission as a function of Galactic latitude and of \hi spin temperature suggests that the DNM is less than 50\% atomic \citep{2011A&A...536A..19P}.
Moreover, doubling the \hi densities in the local ISM because of ubiquitously large \hi opacities would yield a twice lower \g-ray emissivity per gas nucleon, therefore a twice lower CR flux inferred from \g rays in the local ISM. This is clearly at variance with the direct CR measurements performed in the solar system at energies of 0.1 to 1 TeV for which solar modulation is negligible \citep{2015ARA&A..53..199G}.
Large atomic fractions in the DNM being probably the exception for our set of clouds, we have performed the calculations with a 50$\%$ or 100\% molecular composition of the DNM.


Figure \ref{fig:9nhdark} shows the evolution profile of the CO-dark \hd fraction in column density, for a half-molecular DNM composition, as a function of the total gas column density, \nhtot, and of visual extinction, \av, in each of the clouds. The CO-dark diffuse molecular hydrogen exceeds 70$\%$ of the total \hd column density at \nhtot $< 6\times 10^{20}$ cm$^{-2}$, except in the tauM2 region which is centred on the extensively bright part of the Taurus CO cloud. \cite{2010A&A...521L..18V} have observed 53 transition clouds with \hi and $^{12}$CO emissions and no detectable $^{13}$CO emission, thus containing negligible \cosat \hd. In about 30\% of them, the \cii line emission at 1.9 THz indicates CO-dark \hd fractions in column density exceeding 60\%.

The fraction starts to decrease in the 0.4-0.9 magnitude range in visual extinction or in the (0.3-1) $\times$ 10$^{21}$ cm$^{-2}$ range in total gas column density. It then declines with rather comparable slopes in the different clouds. We note that CO emission in the very compact molecular filament of ChaEastI is detected at significantly lower column densities than in the other clouds, possibly because of larger \hd volume densities, but one cannot relate filamentary CO structures to low DNM abundances since the other compact filament in our sample, Musca, retains large DNM column densities up to 2 magnitudes of extinction.

\begin{figure}
  \centering                
  \includegraphics[scale=0.41]{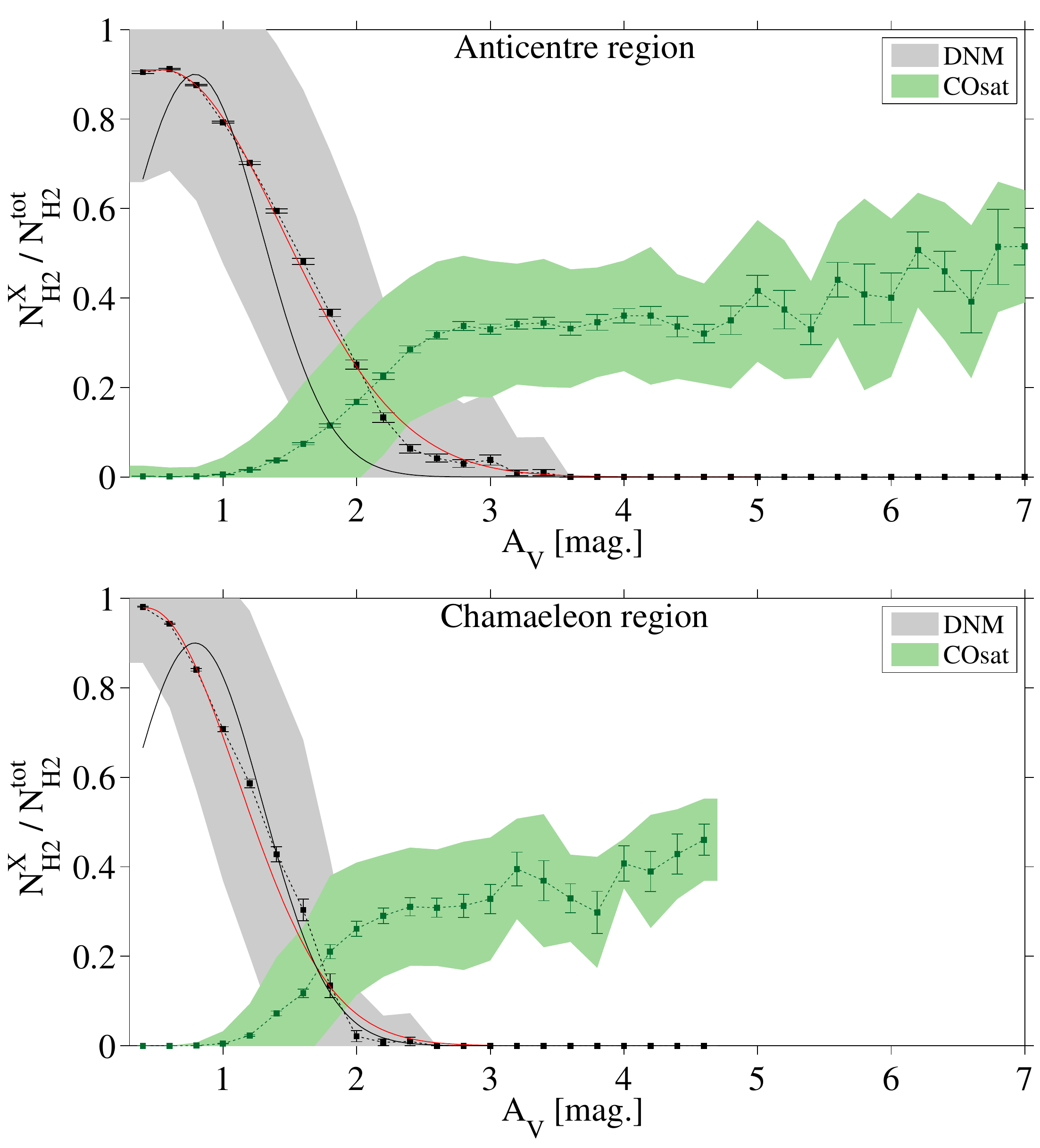}
  \caption{ Evolution with visual extinction \av of the fraction \fcodk of molecular DNM gas (grey) or additional \cosat \hd (green) in the total \hd column-density. The DNM is assumed to be 100\% molecular. The shaded areas and error bars respectively give the standard deviation in the sample and the standard error of the mean. The black line gives the Gaussian profile of \cite{2016ApJ...819...22X} and the red line shows the best fit by a half-Gaussian to the \fcodk fraction.}
   \label{fig:nhAv}
\end{figure}

By integrating the gas over the entire anticentre or Chamaeleon regions, we find the average profiles displayed in Fig. \ref{fig:nhAv} for a fully molecular composition of the DNM to allow comparisons with \cite{2016ApJ...819...22X}. Systematic uncertainties in the current measurements of low extinctions limit our study of the CO-dark \hd fraction to \av > 0.3 magnitude. 
The rms dispersion reflects the differences in individual profiles of Fig. \ref{fig:9nhdark}, due on the one hand to a modest difference, within a factor of 2 to 3, in the amount of extinction required to overcome CO destruction in each cloud and, on the other hand, to the difference in DNM abundance prior to the transition from CO-dark to CO-bright \hd.
On average, 
the CO-dark \hd dominates the column densities up to 1.5 magnitudes in visual extinction. 

The individual and average \fcodk profiles do not compare well with the model prediction of \cite{2010ApJ...716.1191W} that CO emission reaches an optical depth of 1 at \av $\gtrsim 0.7$ magnitude, well beyond the 0.1 magnitude required for the formation of \hd . 
Most of the profiles shown in Figs. \ref{fig:9phasesAV} and \ref{fig:9nhdark} indicate a transition to the CO-bright regime deeper into the clouds, at \av between 1 and 2.5 magnitudes, even though \cite{2010ApJ...716.1191W} have modelled more massive clouds, with a total column density \nhtot of $1.6\times 10^{22}$ cm$^{-2}$, that should screen CO molecules more effectively than the more translucent clouds in our sample. The onset of CO emission is more consistent with the approximate range near 1.5--2.5 magnitudes where the dust optical depth and dust extinction are seen to recover a linear correlation with \hi and CO intensities \citep{2011A&A...536A..19P,2012A&A...543A.103P}. We note that the cloud closest to the model prediction is the very compact filament of ChaEastI, suggesting that the volume density gradient through the cloud plays an important role in addition to the integral screening through the gas columns.

We parametrize the decrease of the CO-dark \hd fraction as a function of \av with a half-Gaussian profile defined as:
\begin{equation}
f_{\rm{COdk \, H_2}}=
\left\lbrace
\begin{array}{ccc}
f_{max}  & \mbox{if} & A_{\rm{V}} \leqslant A_{\rm{V_0}}\\
f_{max} \times \rm{exp}\left[  - \left( \frac{A_{\rm{V}}-A_{\rm{V_0}}}{\sigma_{A_{\rm{V}}}} \right) ^2\right]   & \mbox{if} & A_{\rm{V}} > A_{\rm{V_0}}\\
\end{array}\right.
\end{equation}
The results of the fits for the two analysis regions are given by the red curves in Fig. \ref{fig:nhAv}. We find $f_{max}$=0.91, $A_{\rm{V_0}}$=0.55 mag, $\sigma_{A_{\rm{V}}}$=1.27 mag in the anticentre region and $f_{max}$=0.98, $A_{\rm{V_0}}$=0.43 mag, $\sigma_{A_{\rm{V}}}$=0.97 mag in the Chameleon region. These parameters are derived for a fully molecular DNM, but they change only by a few percent when using a half molecular DNM composition.

\cite{2016ApJ...819...22X} have studied the PDR border of the main Taurus CO cloud in an approximately $0\fdg5$-wide square, centred on $l=172\fdg8$ and $b=-13\fdg2$, at a high angular resolution of 0.05$\degree$. This edge lies within the tauM2 sub-region of our sample. They have observed OH, $^{12}$CO, and $^{13}$CO line emissions and they have inferred the \nhd column densities from the \av extinction observations of \cite{2010ApJ...721..686P}, assuming that the hydrogen is predominately molecular and using the relation \nhd/\av $= 9.4 \times 10^{20}$ cm$^{-2}$ mag$^{-1}$. Their CO-dark \hd fraction is defined as the fraction of molecular gas not traced by $^{12}$CO or $^{13}$CO line emission, so it compares with our measurements of \fcodk in the case of a fully molecular DNM composition. They have modelled the variation of the CO-dark \hd fraction as a function of extinction with a Gaussian profile. 
Given the very different origins of both datasets, the close-up measurement of \cite{2016ApJ...819...22X} agrees remarkably well with the mean evolution of the CO-dark \hd fraction we find in the whole anticentre region. The steeper decline of their CO-dark \hd fraction compared to our mean trend remains compatible with the cloud-to-cloud dispersion in the sample. The discrepancy with the tauM2 profile (Fig. \ref{fig:9nhdark}), however, calls for further investigations in order to understand whether it is due to methodological differences (angular resolution, total gas tracers versus radiative modelling, etc) or to genuine spatial changes in the PDR profiles across tauM2. 

Based on observations of \cii lines at 1.9 THz with \textit{Herschel}, \cite{2014A&A...561A.122L} have estimated the column densities of CO-dark \hd in a large sample of clouds in the Galactic disc and they have measured the CO-dark \hd fraction in the total gas column density. As their estimates of the associated \nhi column densities are very low, their CO-dark \hd fractions in the total gas are close to the \fcodk fraction we have defined in the molecular phase. They have found a very broad range of CO-dark \hd fractions across the whole $10^{21-22}$ cm$^{-2}$ interval. The spread largely encompasses the profiles shown in Fig. \ref{fig:9nhdark} for \nhtot > $10^{21}$ cm$^{-2}$. We note, however, that the transitions we observe in the nearby clouds tend to occur at lower \nhtot column densities than the bulk of the Galactic sample probed with \cii lines. They also occur outside the minimum and maximum fraction bounds predicted by \cite{2009A&A...503..323V} and shown in Fig. 17 of \cite{2014A&A...561A.122L}. Our observations show marked fraction declines at two to three times lower \nhtot column densities than the minimum model prediction, even though the nearby clouds probed in our study have the same solar metallicity and total extinction range (\av < 4 mag) as the gas slabs modelled by \cite{2009A&A...503..323V}. Radiative modelling of those slabs is required to compare the model profiles with our data and with the recorded \wco intensities in order to disentangle the chemical or radiative origin of this discrepancy.

\begin{figure*}[p!]
  \centering                
  \includegraphics[scale=0.585]{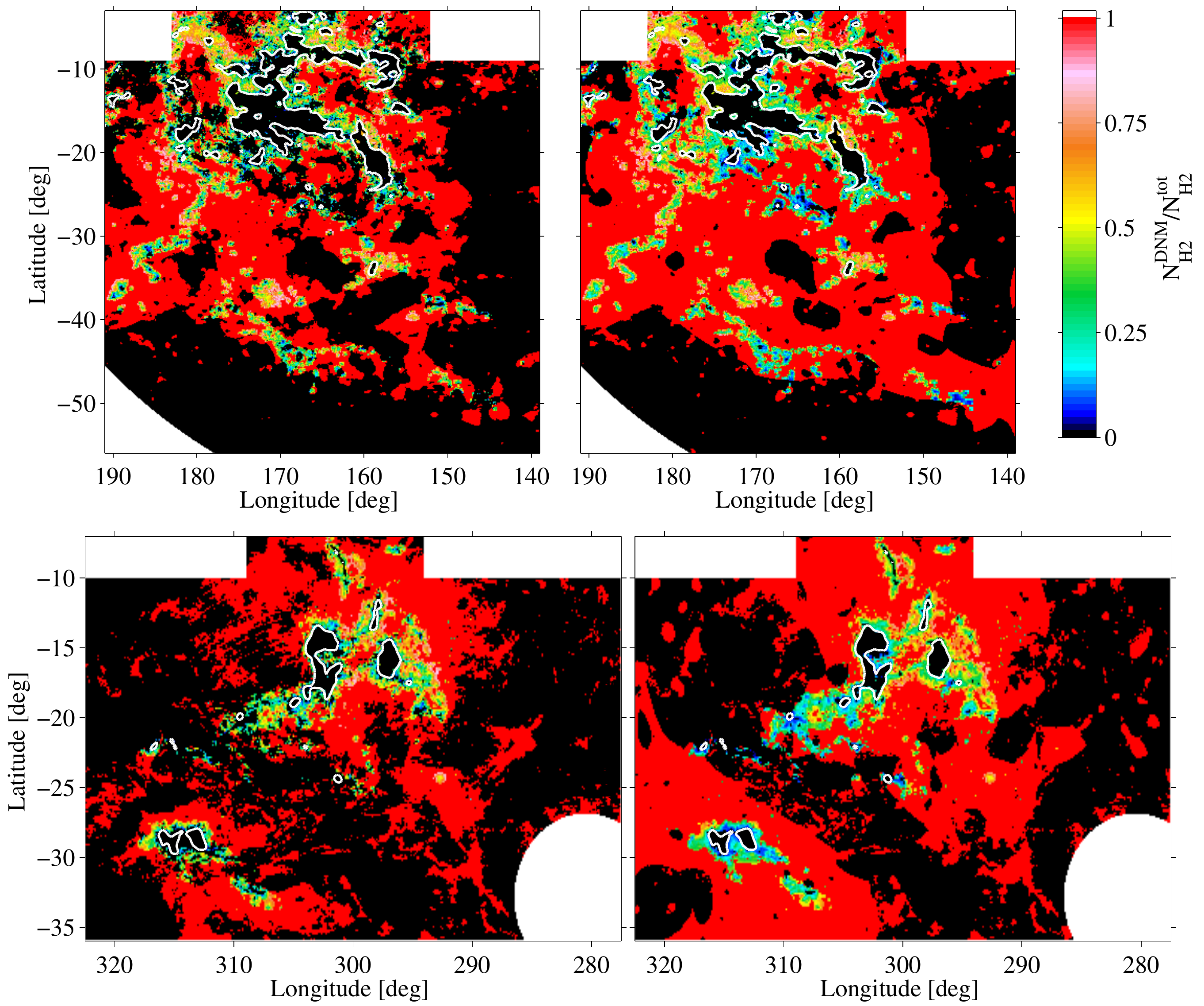}
  \caption{ Fraction of the molecular DNM in the total \hd column-density (\fcodk) derived from the dust (left) and \g-ray (right) analyses for the anticentre (top) and Chamaeleon (bottom) regions. The DNM is assumed to be $50\%$ molecular. The white contours outline the shapes of the CO clouds at the 7 \wcounit level chosen to separate DNM and \cosat components.}
   \label{fig:fdgmap}
\end{figure*}

\input{SubMass.txt}

Figure \ref{fig:fdgmap} illustrates how the CO-dark \hd fraction relates to the structure of the molecular clouds. Generally \fcodk progressively decreases from the DNM outskirts to the CO-bright edges, but local fluctuations imply a more complex pattern that is difficult to interpret in the 2D projection. Extended or elongated DNM-rich structures are more susceptible to photo-dissociation along the short axes of the cloud.
The transition from the atomic to molecular phase is expected to occur at different column densities according to the strength of the ambient ISRF, outside and inside the cloud, according to the 3D spatial distribution of the gas and shadowing effects, and according to the 3D structure in density which impacts the ambient \hd and CO formation rates. 
Theoretical predictions by \cite{2009ApJ...693..216K} indicate that the  fraction of molecular gas in a galaxy depends primarily on its column density and is, to a good approximation, independent of the strength of the interstellar radiation field. According to the spherical PDR model of \cite{2010ApJ...716.1191W}, an increased ISRF intensity implies a modest diminution in the thickness of the diffuse \hd envelope surrounding the CO cloud, but the ratio of \hd and CO depths remains constant, so the CO-dark \hd fraction remains stable against changes in ISRF intensity. Yet, this stability depends on the choice of volume density profile inside the cloud ($n_{\rm H} \propto 1/r$). For this profile they find only 30\% changes in the CO-dark \hd fraction for UV radiation fields ranging between 3 and 30 times larger than the local value of \cite{1978ApJS...36..595D}.


We do not have a direct observational measure of the strength of the ISRF, but the specific power radiated by large dust grains per gas nucleon provides an indirect estimate of their heating rate by stellar radiation if the grains have reached thermal equilibrium. We have used the radiance map of \cite{2014A&A...571A..11P} and our distributions of \nhtot column densities to map the specific power of the grains across the anticentre and Chamaeleon regions. 
We find only a 20$\%$ dispersion around the mean value of the specific power in the $0.4 <$ \av$<1.5$ magnitude range that is critical for the {H\,{\sc i}-\hd transition. We find no evidence of a spatial correlation between the distributions of the grain specific powers and those of the CO-dark \hd fractions shown in Fig. \ref{fig:fdgmap}.
As the strength of the ISRF potentially plays a marginal role in the relative contributions of CO-dark and CO-bright \hd, the density structure of a cloud in 3D and the internal level of turbulence may be the key parameters to explain the small spread (by a factor of 4) in \nhtot and \av thresholds required to overcome CO destruction and sustain a population of CO molecules dense enough to be detected by the current CO surveys. Simulation of \hd formation within dynamically evolving turbulent molecular clouds shows that \hd is rapidly formed in dense clumps and dispersed in low-density and warm media due to the complex structure of molecular clouds \citep{2016A&A...587A..76V}.
 
We now turn to the transition from CO-bright to CO-saturated \hd that occurs in the cores of CO clouds when the optical thickness to $^{12}$CO emission is large enough.  Figure \ref{fig:9phases} indicates that the CO-bright part of the total hydrogen column density saturates at \av $\sim 2$ magnitudes whereas the \cosat fraction still increases.
We follow the variation of the \cosat fractions in the total \hd column-density in Fig. \ref{fig:nhAv}.
They reach 40 $\%$ at \av$>4$ magnitude and 50$\%$ at \av $> 6$ magnitude. As mentioned in Sec \ref{sec:res}, our analysis cannot separate the additional \cosat gas from the foreground and background DNM present along the lines of sight towards the dense regions of large \wco intensities, i.e., large \av extinctions. The interpolation of the DNM \hd column densities across the \cosat regions indicates that the \cosat \hd fractions could be lowered by about 5-10$\%$ after subtraction of the DNM contributions.


\subsection{ Average CO-dark \hd fractions in mass and properties of the clouds }
\label{sec:meanFrac}

This section focuses on the variations of the average CO-dark \hd fractions in mass from cloud-to-cloud.
We have integrated the gas masses over the extent of each cloud in each of the {H\,{\sc i}}, DNM, CO, and \cosat phases.
The masses are directly derived from the column density maps. All quoted masses include the Helium contribution.
We have used the column density maps derived from the \g-ray analyses, which are more robust against dust evolution (see details of their derivation in Sec \ref{sec:coldens}). The distance used for each cloud is the same as that of its respective parent complex given in Table \ref{tab:CLmass}. The mass ratios are not affected by distance uncertainties. Table \ref{tab:9mass} lists the relative contributions of the different phases to the total gas mass of clouds. The number distribution for each gas phase is shown in Fig. \ref{fig:hst_frac}.
We have also calculated the CO-dark \hd fraction in mass, integrated over each cloud, as:
\begin{align}
F_{\rm{COdk \, H_2}}= \frac{M_{\rm{H_2}}^{\rm{DNM}}}{M_{\rm{H_2}}^{\rm{tot}}}=\frac{M_{\rm{H_2}}^{\rm{DNM}}}{M_{\rm{H_2}}^{\rm{DNM}}+M_{\rm{H_2}}^{\rm{CO}}+M_{\rm{H_2}}^{\rm{COsat}}}
\label{eq:defFcodk}
\end{align}
Table \ref{tab:9mass} provides the results for 50\% and 100\% molecular DNM composition.

Figure \ref{fig:hst_frac} highlights that atomic hydrogen is the dominant form of gas by mass in all the sampled clouds. The DNM is the second most important contributor, with mass fractions often close to 20\%, in close agreement with the average of 25\% found in the sample of 53 clouds exhibiting CO-dark \hd excesses in \cii line emission \citep{2010A&A...521L..18V}.  Bright CO clouds such as the star-forming tauM2 region or the compact ChaEastI filament gather only $\sim$10\% of their mass at the DNM interface. On the contrary, the fraction increases up to 27$\%$ in faint CO clouds such as tauS1 and tauS2.  The values in Table \ref{tab:9mass} show that the relative amount of molecular gas present in the CO-dark and CO-bright regions varies greatly from cloud to cloud. Generally, the bright CO clouds, with a mean \wco intensity above 4 \wcounit or a maximum intensity above 20 \wcounit, have more mass in the CO-bright region than in the DNM envelopes, contrary to the CO-faint and rather diffuse clouds. This is reflected by the average CO-dark \hd fractions varying from 10$\%$ to 50$\%$ in the brighter molecular clouds (tauM, Aur, Cal, Cha) and between 50$\%$ and 90$\%$ in the diffuse clouds (Cet, tauS, tauN). 

\citet{2015MNRAS.448.2187C} have studied a wider region of the Galactic anticentre that includes the clouds studied here and the Orion clouds, at an angular resolution of 0\fdg1. They have modelled the stellar extinction data as a linear combination of \hi column densities from the GALFA survey and CO intensities from the \textit{Planck} type 3 map, corrected for $^{13}$CO contamination by multiplying the map by 0.86 \citep{2014A&A...571A..13P}. \citet{2015MNRAS.448.2187C} used the differences between the data and model to derive the column density of DNM gas and then calculate the CO-dark \hd mass fraction. They found this mass fraction to be 24$\%$ in Taurus, 31$\%$ in Orion, and 47$\%$ in Perseus. For the high-latitude clouds TauE2 and TauE3, which approximately correspond to the tauS2 and tauS1 clouds of our sample, they found values of 80$\%$ and 77$\%$. For a 50-100$\%$ molecular composition of the DNM, we respectively find CO-dark \hd mass fractions of 11--19$\%$ in Taurus (tauM2), 0.82--0.90$\%$ in tauS2, and 0.79--0.88$\%$ in tauS1, in close agreement with the results of \citet{2015MNRAS.448.2187C}. 


 \begin{figure}
  \centering                
  \includegraphics[width=\hsize]{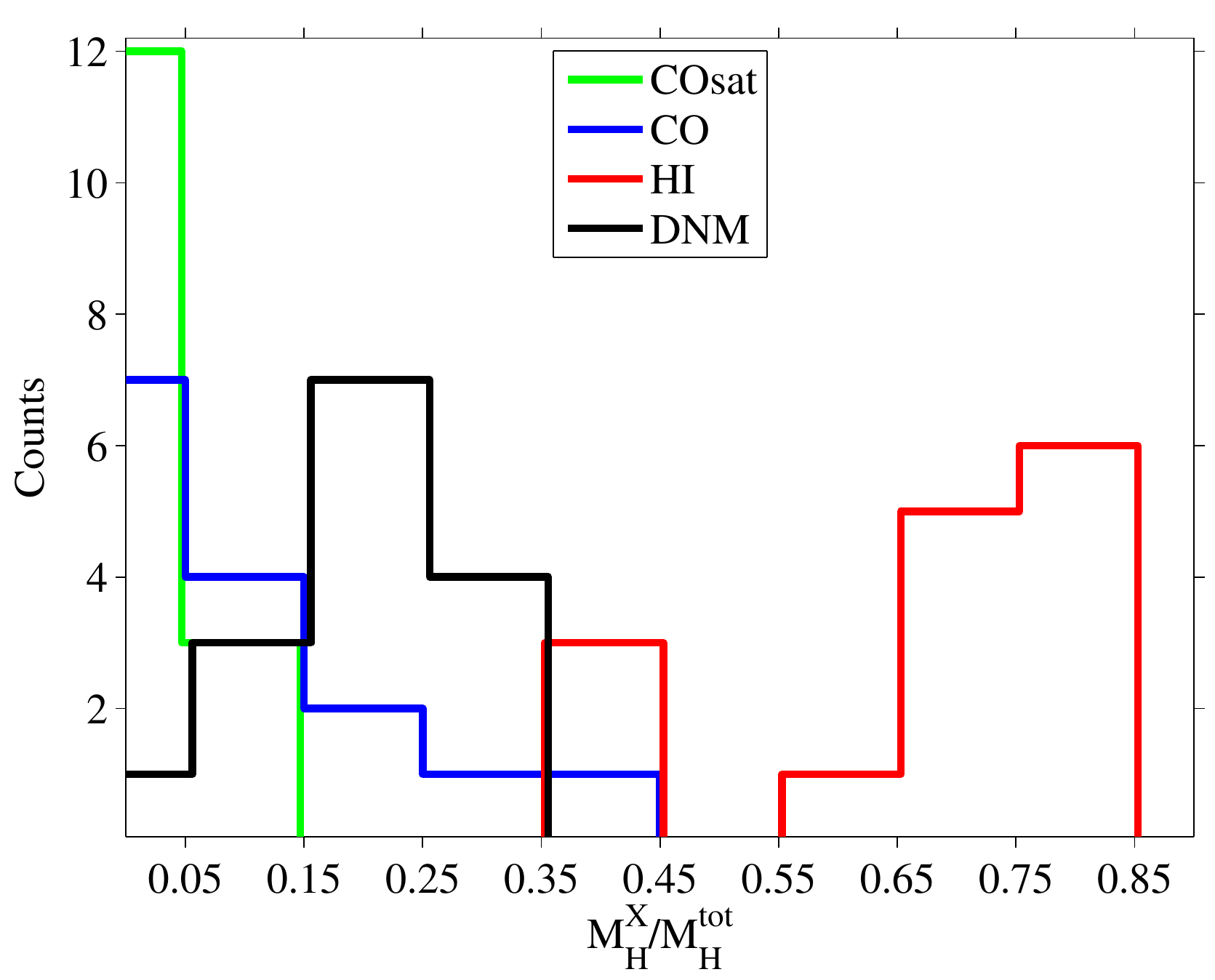}
  \caption{ Number distributions of the HI, DNM, CO, and \cosat fractions of the total mass found in the sample of clouds for each phase.}
   \label{fig:hst_frac}
\end{figure}

\begin{figure*}
  \centering                
  \includegraphics[width=\hsize]{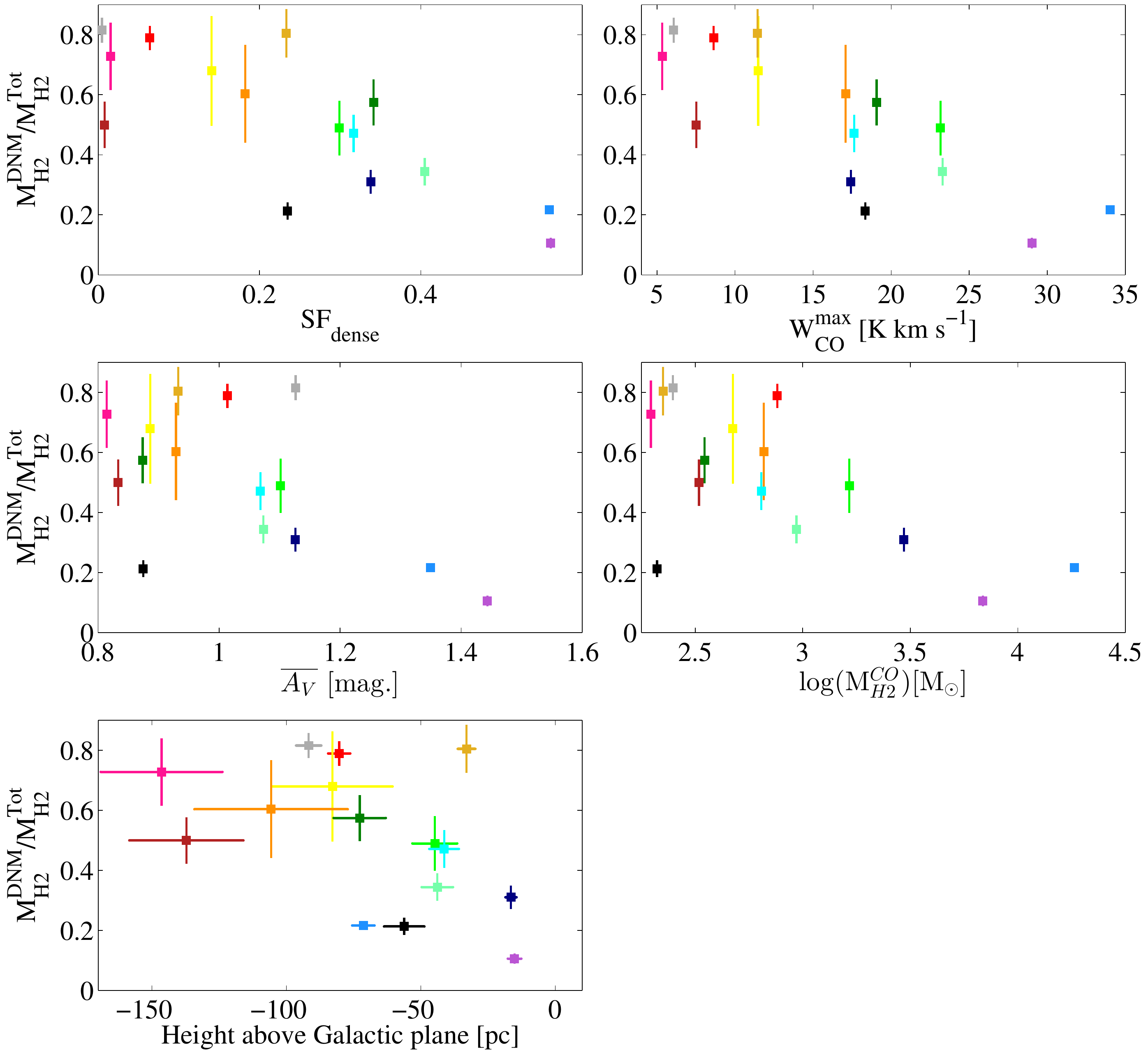} 
  \caption{ Average fraction of the molecular DNM in the total H$_{2}$ mass (\fcodkm) as a function of: the surface fraction of dense gas SF$_{\rm{dense}}$ in the cloud (top left), the maximum CO integrated intensity $W_{\rm{CO}}^{\rm{max}}$ recorded in the cloud (top right), the average visual extinction $\overline{A_{\rm{V}}}$ over the cloud (middle left), the logarithm of the \hd mass in the CO phase $M_{\rm{H_2}}^{\rm{CO}}$ (middle right), and the height above the Galactic plane (bottom left). The DNM is assumed to be $50\%$ molecular. The colors refer to the same clouds as in Fig. \ref{fig:subRegions}. }
   \label{fig:fdg_moy}
\end{figure*}

\begin{figure}
  \centering                
  \includegraphics[width=\hsize]{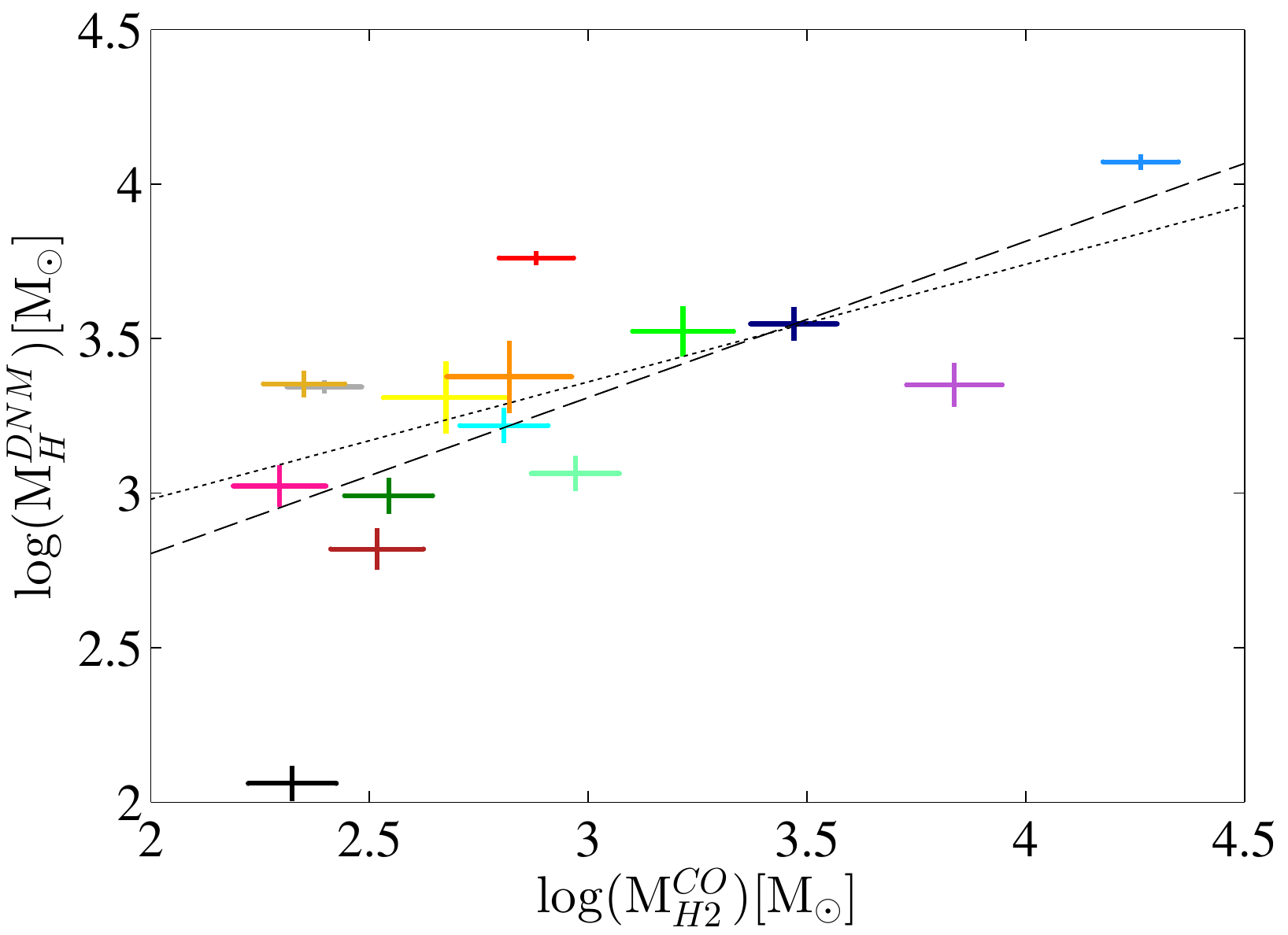} 
  \caption{ Gas mass in the DNM (both atomic and molecular) as a function of the H$_{2}$ mass in the CO-bright phase. The dashed and dotted lines correspond to the best power-law fit respectively including and excluding the compact ChaEastI filament (black point). The colors refer to the same clouds as in Fig. \ref{fig:subRegions}. }
   \label{fig:logMass}
\end{figure}

We further discuss the relative contribution of the diffuse CO-dark \hd to the total molecular mass according to the cloud properties and location.
Figure \ref{fig:fdg_moy} shows the evolution of the CO-dark \hd mass fraction, \fcodkm, as a function of: 
\begin{itemize}
\item the surface fraction in solid angle of dense regions with large \wco intensity inside a cloud, $\rm{SF}_{dense}=\frac{S(W_{\rm{CO}>5 \rm{K km s^{-1}}})}{S(W_{\rm{CO}>1 \rm{K km s^{-1}}})}$; this parameter reflects the compactness of the molecular gas;
\item the maximum CO intensity, $W_{\rm{CO}}^{\rm{max}}$, recorded in the cloud; 
\item the average visual extinction, $\overline{A_{\rm{V}}}$, of the cloud; 
\item the \hd mass found in the CO-bright phase (excluding the \cosat component), $M_{\rm{H_2}}^{\rm{CO}}$;
\item the cloud height above the Galactic plane;
\end{itemize}

We find a significant trend for \fcodkm to decrease from 0.8 to 0.1 with increasing  $\rm{SF}_{dense}$ and $W_{\rm{CO}}^{\rm{max}}$. A negative slope is detected with a confidence level larger than 18 $\sigma$ for these diagnostics, indicating that the fraction tends to decrease from diffuse to more compact clouds. Figure \ref{fig:fdg_moy} shows that the diffuse clouds, rich in CO-dark \hd, also tend to lie at larger heights above the Galactic plane than the brighter and more compact CO clouds. There is a possible bias due to velocity crowding and an increased difficulty in separating the different complexes and gas phases in the dust and gamma-ray models as the latitude decreases. We note, however, that the DNM-rich cloud of Musca lies close to the plane. Conversely, the medium-latitude clouds of TauM2, ChaII-III, and ChaEastI, well separated from the Galactic disc background, are DNM poor.

Using the thermal emission of dust grains over the whole sky at $|b| >5\degree$ and masking out CO regions with \wco intensities larger than 1 K km s$^{-1}$, \citet{2011A&A...536A..19P} have found an average \fcodkm mass fraction of $0.54$ close to the value of 0.62 obtained at $|b| > 10\degree$ from dust visual extinction \citep{2012A&A...543A.103P}. These high averages indeed reflect the local predominance of molecular clouds with moderate masses and diffuse structures in the sample of nearby clouds seen at medium latitudes. \citet{2016ApJ...833..278M} have studied several nearby molecular clouds at intermediate latitudes, namely MBM 53, 54, and 55 and the Pegasus loop. They have used the dust optical depth and the dust radiance, derived from \textit{Planck} and IRAS data, in combination with \g-ray observations from \textit{Fermi} LAT in order to monitor the variations of the dust emission properties as a function of the dust temperature, and to estimate the total gas column density. They have found that the gas mass in the DNM, not traced by \hi and CO surveys, is up to 5 times larger than the molecular gas mass traced by CO emission. So according to the definition in Eq. \ref{eq:defFcodk}, their CO-dark \hd fraction reaches up to 75--83\% for a 50--100\%  molecular DNM composition. Although their method differs from ours, this result agrees with the fraction we have found in similar translucent clouds of the anticenter region (Cet, tauN, tauS in Table \ref{tab:9mass}).

While it is clear in Fig. \ref{fig:fdg_moy} that obscured clouds with $\overline{A_{\rm{V}}}$ in excess of 1.2 mag have low CO-dark \hd mass fractions, below 20\%, the large scatter at low $\overline{A_{\rm{V}}}$ indicates that this observable cannot be used to decide whether the CO-dark \hd dominates over the CO-bright component in the molecular census. For this purpose, the peak CO intensity in a cloud may turn out to be more useful. Adding other clouds to the sample will confirm if clouds with $W_{\rm{CO}}^{\rm{max}}$ less than approximately 20 \wcounit have more molecular hydrogen in the DNM than in the CO-bright parts. The model of \cite{2010ApJ...716.1191W} predicts that, for a given cloud, \fcodkm is a strong decreasing function of the mean extinction of the cloud, $\overline{A_{\rm{V}}}$. Yet, their modelled fractions remain very large, near 75\%, at $\overline{A_{\rm{V}}} = 2$ magnitudes, at variance with the low fractions we find already above $\sim$1.2 magnitude.


Except for the outlier CO filament of ChaEastI, compact and filamentary, which has particularly low mass, we find a significant decrease of the \fcodkm fraction with the logarithm of the $M_{\rm{H_2}}^{\rm{CO}}$ mass. The negative slope is preferred relative to a constant at a confidence level exceeding 19 $\sigma$. Figure \ref{fig:logMass} further shows that the mass present in the DNM envelopes approximately scales with the square root of the molecular mass seen in CO. We find a correlation coefficient of 0.86 between the two sets of masses on a logarithmic scale.  A power-law fit yields:
\begin{equation}
M_{\rm H}^{\rm DNM} = (62 \pm 7) {M_{\rm H_2}^{\rm CO}}^{\,0.51 \pm 0.02}
\label{eq:Mdnm_Mco_scaling}
\end{equation}
 if we include all the clouds, and
 \begin{equation}
 M_{\rm H}^{\rm DNM} = (166 \pm 19) {M_{\rm H_2}^{\rm CO}}^{\,0.38 \pm 0.02}
\end{equation}
 if we exclude the outlier filament of ChaEastI.  
 
The slopes are detected at the 13~$\sigma$ and 24~$\sigma$ confidence levels, respectively. They compare well with the DNM mass evolution, $M_{\rm{H}}^{\rm{DNM}} \propto {M_{\rm{H_2}}^{\rm{CO}}}^{\,0.4}$, found at latitudes $|b| > 5\degree$ in broader and more massive molecular complexes than the clouds studied here \citep{2005Sci...307.1292G}. The former relation had been derived with a coarser derivation of the dust optical depth (from \textit{IRAS} and \textit{DIRBE}) and lower-resolution data in \g rays (from \textit{EGRET}) and in \hi lines \citep[LAB survey, ][]{2005A&A...440..775K}. The data points in Fig. \ref{fig:logMass} strengthen the reliability of a power-law relation between  the DNM mass and the CO-bright \hd component. They also extend its application to lower CO masses. According to Larson's laws, gravitationally bound molecular clouds have masses that approximately scale with the square of the cloud linear size \citep{1981MNRAS.194..809L,2010A&A...519L...7L}. For bound clouds, the scaling relation of equation \ref{eq:Mdnm_Mco_scaling} implies that the mass present in the DNM envelopes should scale with the linear size of the molecular clouds. This can be observationally tested by using the kinematical information of the \hi and CO lines when firm distances to these clouds can be inferred from the stellar reddening information from \textit{Gaia}.
This relation between masses in the DNM and CO phases can also serve to extrapolate the amount of DNM gas beyond the local ISM and in external galaxies. Its confirmation is of prime importance, but it requires further tests in giant molecular clouds of larger CO masses than in Fig. \ref{fig:logMass}, in compact clouds of small size, and in clouds of different metallicity. 

The PDR model of \citet{2012A&A...544A..22L} yields a mass fraction of 32\% in a cloudlet with only 9.5 M$_\odot$ in CO and illuminated by a standard UV field. The model of \cite{2010ApJ...716.1191W} yields a narrow range of fractions between 25\% and 33\% for giant molecular clouds with masses of $(0.1-3)\times10^6$ M$_\odot$ much larger than those probed in our sample and with a UV radiation field 3 to 30 times larger than the local value of \cite{1978ApJS...36..595D}. The theoretical studies therefore point to rather stable CO-dark \hd fractions that should little depend on cloud mass or radiation field. Our results challenge this view by showing a wide range of mass fractions that relate to cloud properties such as the CO diffuseness ($\rm{SF}_{dense}$) and the CO brightness (peak \wco intensity or integral of \wco intensities that gives the $M_{\rm{H_2}}^{\rm{CO}}$ mass). 

Cloud-dependent variations call for care in the interpretation of averages taken 
over large ensembles of clouds with different evolutionary or structural properties. For instance, the difference noted between the local clouds seen toward the inner and outer Galaxy by \citet{2012A&A...543A.103P} does not relate to the Galactocentric distance, but to different combinations of clouds in different states. This could explain the discrepancy noted by \cite{2014MNRAS.441.1628S} between an `apparent  decrease' of \fcodkm with Galactocentric distance in \cite{2012A&A...543A.103P} and an increase with this distance in the much broader sample of clouds studied with \cii lines across the Milky Way \citep{2013A&A...554A.103P}. Moreover, the present trend for larger \fcodkm fractions in clouds lying higher above the Galactic plane, if confirmed, should also raise concerns for radial studies across the Galactic disc in constant latitude bands. The rise observed in \fcodkm from about 10\% at 4 kpc up to 60--80\% near 10 kpc is not biased by such height variations because it is based on \cii line measurements taken along the Galactic plane, at $b=0\degree$ \citep{2013A&A...554A.103P}.


In their simulation of the Milky Way, \cite{2014MNRAS.441.1628S} use a uniform IRSF, metallicity, and mean gas surface density, so they cannot reproduce a global variation of \fcodkm with Galactocentric radius. In conditions akin to those in the local ISM, they obtain an average fraction of 42\% that is almost insensitive to the \wco intensity threshold used to separate CO-bright and CO-dark gas. They have, however, found notable differences in \fcodkm between the arm and inter-arm regions. Generally the fraction is anti-correlated with the total molecular \nhd column density. The molecular gas is predominantly CO-bright in the spiral arms ($0.1 <$\fcodkm$<0.5$), but the CO-dark \hd mass fraction exceeds 80\% in the inter-arm regions where the clouds are sheared out by the differential rotation of the disc following their passage through the spiral arm. The stretched cloud structures become more permeable to dissociating radiation.
Interestingly, we find in our sample a wide range of CO-dark \hd mass fractions responding to properties such as $\rm{SF}_{dense}$, $M_{\rm{H_2}}^{\rm{CO}}$, and Galactic height, that can reflect the transitory/sheared versus gravitationally well-bound state of a cloud. 

Figure \ref{fig:WCOvsNH2} shows the distribution of \wco intensities as a function of \nhd column densities observed in the anticentre and Chamaeleon regions. The data points above the grey line in this figure correspond to directions where \xco<$10^{20}$ cm$^{-2}$K$^{-1}$km$^{-1}$s (as in the Chamaeleon, main Taurus, California, and Perseus clouds). The data points gathering along the horizontal axis, at null or very low \wco up to large \nhd values, correspond to the DNM gas. 
In the simulation of \cite{2014MNRAS.441.1628S} with local-ISM characteristics, CO molecules predominantly form in spiral-arm clouds, which emit mostly at \nhd$>10^{21}$ cm$^{-2}$  with \wco$>10$ \wcounit as shown by the blue and red contours in Fig. \ref{fig:WCOvsNH2}.
The comparison highlights that the simulation produces underluminous CO clouds at low column densities compared to the observations. This was also noted by \citet{2012AeA...544A..22L}. The CO deficit in their simulation reached a factor of ten in column density in the peripheral regions exposed to the UV radiation, where \nhd $\lesssim 2\times 10^{20}$ cm\msq. So, there seems to be a general problem with our understanding of chemistry at low gas density in the photo-dominated regions (PDR), either in the treatment of the UV attenuation or in the absence of warm chemistry driven by intermittent energy deposition in regions where the interstellar turbulence is dissipated \citep{2014AeA...570A..27G}. This under-prediction of CO emission biases the simulated values of CO-dark \hd fractions upward in column density and in mass. Additionally simulations do not predict bright CO emission at larger \nhd column densities of a few $10^{21}$ cm$^{-2}$, as observed in nearby molecular clouds. This explains why simulations obtain twice larger \xco factors than the values measured in several nearby clouds. We have discussed this discrepancy in \cite{2017A&A...601A..78R}. These specifics of \cite{2014MNRAS.441.1628S} simulation explains why the simulated CO emission gathers primarily in giant molecular clouds in the spiral arms and why the simulations yield a sharp contrast in CO-dark \hd abundance inside and outside of the arms.



 \begin{figure}
  \centering                
  \includegraphics[width=\hsize]{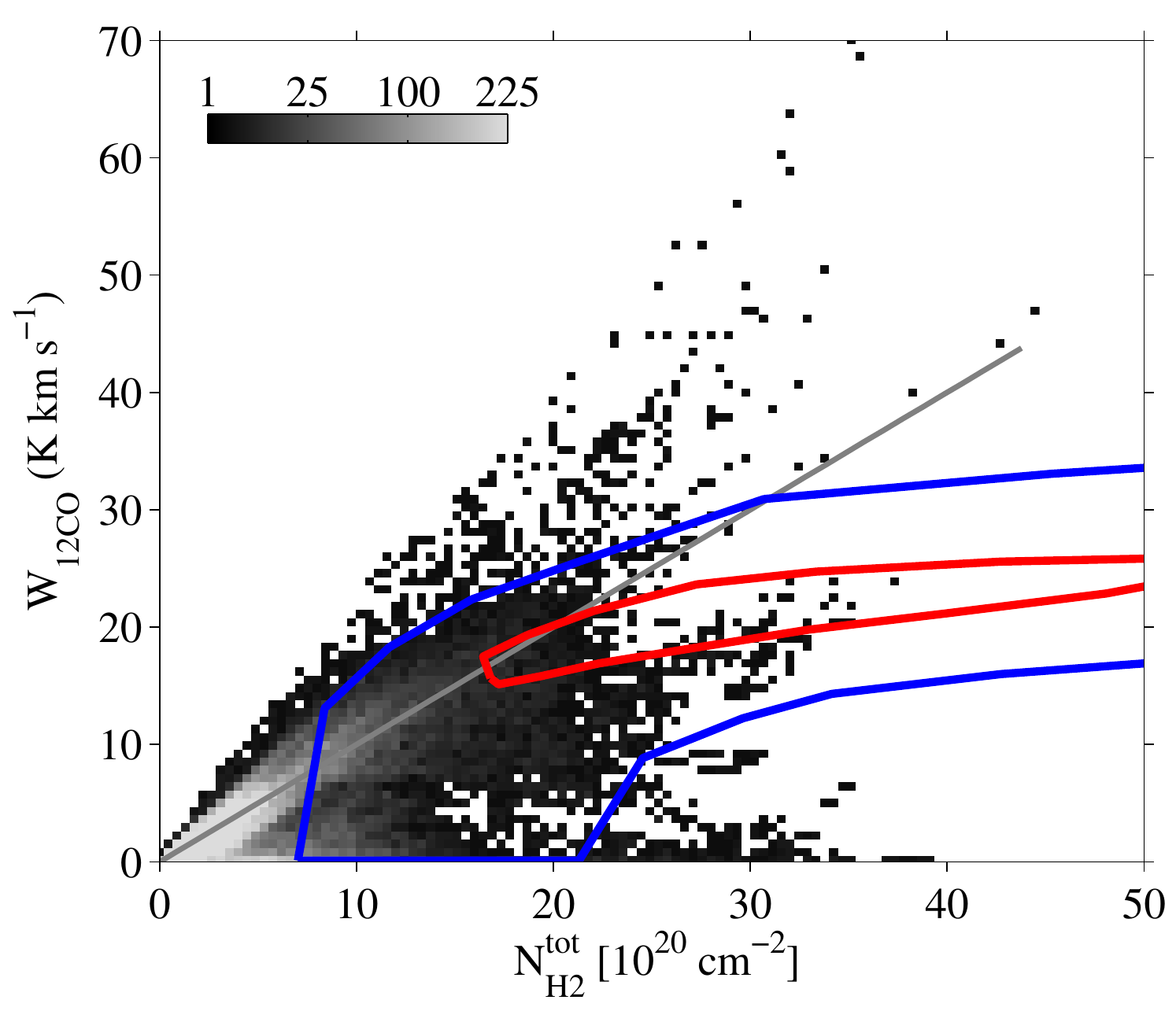}
  \caption{ Number distribution of the integrated $^{12}$CO intensity, \wco, as a function of the total H$_{2}$ column-density derived from the \g-ray fits for the anticentre and Chamaeleon regions. In each direction \nhdtot is the sum of $N_{\rm{H_2}}^{\rm{DNM}}$, $N_{\rm{H_2}}^{\rm{CO}}$, and $N_{\rm{H_2}}^{\rm{COsat}}$ contributions. The DNM is assumed to be $50\%$ molecular. We use the \xco factors derived from \g rays to estimate $N_{\rm{H_2}}^{\rm{CO}}$.  The grey line is an arbitrary reference corresponding to \xco = \xcounit. The blue and red contours delimit the least and most populated parts of the distribution found in a simulation of the Milky Way by \cite{2014MNRAS.441.1628S}.}
   \label{fig:WCOvsNH2}
\end{figure}
\section{Conclusions}

We have analysed the gas, dust, and CR content of several nearby clouds in the anticentre and Chamaeleon regions. As described in \cite{2015A&A...582A..31A} and \cite{2017A&A...601A..78R}, in order to trace the total gas column density, we have modelled the \g-ray emission and dust optical depth at 353 GHz  as linear combinations of several gaseous components associated with different phases (ionized, {H\,{\sc i}}, DNM, CO, and \cosat media). We have selected a subset of 15 clouds with no overlap in direction to avoid confusion between the separate phases of the various clouds. Thereby, we have been able to follow the relative contributions of the different gas phases to the total mass and column density, and to study phase transitions in the clouds. The main results are summarized as follows:
\begin{itemize}
\item \textbf{On the gas content:}
\hi is the dominant form of gas by mass in our sample. It contributes 50$\%$ to 80$\%$ to the total mass. The capability of CO lines to trace the total \hd gas varies widely from cloud to cloud. The DNM envelopes, composed of dense optically-thick \hi and diffuse CO-dark \hd at the transition between the atomic and molecular phases, contribute about 20$\%$ to the cloud masses in our sample. The DNM mass often exceeds the \hd mass seen in CO in diffuse clouds (see below) and it approximately scales with the square root of the CO-bright mass.
The CO-saturated \hd gas, which is seen in addition to the \wco-inferred component in dense regions of large optical thickness to  $^{12}$CO lines, generally contributes less than 10\% to the total cloud mass. It accounts for 40$\%$ of the molecular column densities at \av>3 magnitudes.
\item \textbf{On phase transitions:} The transition from the \hi to \hd composition and the onset of visible CO emission both occur in narrow intervals of total gas column density in the selected clouds, respectively at $(0.6-1) \times 10^{21}$ cm$^{-2}$ and $(0.6-2.5) \times 10^{21}$ cm$^{-2}$. We observed a systematic decrease in the \hi fraction in the total column density  above \nh$\approx10^{21}$ cm$^{-2}$ (\av$\approx 0.4$ magnitude). The CO intensities rise in the 1-2.5 magnitude range. The $^{12}$CO emission significantly saturates at \av>3 magnitudes. 
\item \textbf{On the CO-dark \hd fractions:}   The \fcodk fraction in column density is observed to decrease from the edge to the core of an individual cloud or a broader complex. The \fcodk decrease with increasing \av follows a half-Gaussian profile peaking at 0.4--0.6 magnitude and with a standard deviation close to 1 magnitude. The dust specific power reflecting the strength of the ISRF varies only marginally across the DNM envelopes and it does not correlate with spatial changes in \fcodk. In the solar neighbourhood, differences in the \fcodkm mass fraction primarily stem from the cloud structure: \fcodkm significantly decreases from diffuse to dense molecular clouds. The DNM molecular mass often exceeds the \hd mass seen in CO for clouds with peak $W_{\rm{CO}}^{\rm{max}}$ intensities lower than about 15-20 K km s$^{-1}$,  with surface fractions of bright CO intensities $\rm{SF}_{dense}$ less than about 0.3, or with average visual extinctions $\overline{A_{\rm{V}}}$ lower than about 1 magnitude. We note a trend for larger \fcodkm fractions in clouds located at higher latitudes above the Galactic plane.
\end{itemize}

In future studies, it will be interesting to compare maps of the total gas derived jointly from \g rays and dust optical depth, or dust extinction, to the gas traced by other species. In diffuse environments, CO precursors such as CH \citep{2010A&A...521L..16G}, OH \citep{2015AJ....149..123A}, and HCO \citep{2014A&A...564A..64L} can probe the CO dark \hd in the DNM with less confusion than \cii lines. In dense molecular environments, CO isotopologues such as $^{13}$CO and C$^{18}$O, and rarer molecules routinely probe the CO-saturated molecular gas at large volume densities. 
Observations of these tracers are, however, still sparse and their angular extent is often inadequate for comparisons with gas maps inferred from \g-ray and dust studies. Dedicated campaigns toward individual nearby clouds are required to fully explore the relative merits and limitations of the different tracers and to constrain the DNM composition.

Additionally, extending our sample to other clouds in the local ISM, with other masses and various heights above the Galactic plane, then at larger distance in the Local Arm and in between the Local and Perseus arms, is necessary to better characterize the trends in \fcodkm discussed here. The results would greatly improve our view of the relative distributions of CO-dark and CO-bright \hd gas across the Galaxy.

\begin{acknowledgements}
The \textit{Fermi} LAT Collaboration acknowledges generous ongoing support
from a number of agencies and institutes that have supported both the
development and the operation of the LAT as well as scientific data analysis.
These include the National Aeronautics and Space Administration and the
Department of Energy in the United States, the Commissariat \`a l'Energie Atomique
and the Centre National de la Recherche Scientifique / Institut National de Physique
Nucl\'eaire et de Physique des Particules in France, the Agenzia Spaziale Italiana
and the Istituto Nazionale di Fisica Nucleare in Italy, the Ministry of Education,
Culture, Sports, Science and Technology (MEXT), High Energy Accelerator Research
Organization (KEK) and Japan Aerospace Exploration Agency (JAXA) in Japan, and
the K.~A.~Wallenberg Foundation, the Swedish Research Council and the
Swedish National Space Board in Sweden. Additional support for science analysis during the operations phase is gratefully acknowledged from the Istituto Nazionale di Astrofisica in Italy and the Centre National d'\'Etudes Spatiales in France.
The authors acknowledge the support of the Agence Nationale de la Recherche (ANR) under award number STILISM ANR-12-BS05-0016. 
\end{acknowledgements}

\bibliographystyle{aa}
\bibliography{biblioG2}


\end{document}

%% file: DistMass_IG.txt
\renewcommand{\arraystretch}{1.25}
\setlength{\tabcolsep}{0.19cm}
\begin{table*}
\caption{Distance to the clouds, mass in their \hi and CO phases, SFR and \xco factors}
\centering
\begin{tabular}{l|c|c|c|c|c|c}
 & Distance &  M$_{\rm{HI}}$  & M$\rm{_{H_2}^{CO}}$ & M$\rm{_{H_2}^{CO}}$ ($A_{\rm V}$>8 mag) & SFR & X$_{CO}^{\gamma}$\\ 
Name & [pc]& [M$\odot$]  & [M$\odot$] & [M$\odot$] & [10$^{-6}$ M$_{\odot}$ yr$^{-1}$]& $10^{20}$cm$^{-2}$K$^{-1}$km$^{-1}$s\\ 
\hline 
\textbf{{\color{deepPink} Cet1}} &  190   $\pm$30& 4640  $\pm$720 &200 $\pm$50 & -&  -& 1.01$\pm$0.16\\ 
\textbf{{\color{firebrick} Cet2}} & $"$ & 3720  $\pm$580 &330  $\pm$80 & -&- & " \\ 
\textbf{{\color{lime} tauM1}} & 140   $\pm$30& 11500 $\pm$2100&1650  $\pm$440 &- & - & 0.67$\pm$0.01\\ 
\textbf{{\color{mediumOrchid} tauM2}} & $"$& 6920 $\pm$1150 &6870  $\pm$1730 &410 $\pm$100 &18.5  $\pm$ 10.3 &  $"$ \\ 
\textbf{{\color{red} tauS1}} & 160   $\pm$10 & 14800  $\pm$800 &760  $\pm$150 &- & -& 1.04$\pm$0.05 \\ 
\textbf{{\color{darkGrey} tauS2}} & $"$& 6100 $\pm$300 &250  $\pm$50 &- & -& $"$ \\ 
\textbf{{\color{yellow} tauN1}} & 190   $\pm$50 & 8100 $\pm$2200 &470  $\pm$150 &- &- & 0.90$\pm$0.03 \\ 
\textbf{{\color{darkOrange} tauN2}} &$"$ & 10600  $\pm$2900 &660  $\pm$220 &27 $\pm$9 &1.2  $\pm$ 0.7 & $"$ \\ 
\textbf{{\color{navy} Auriga}}  & $"$& 5290 $\pm$670 &2950  $\pm$670 & -& - & $"$ \\ 
\textbf{{\color{dodgerBlue} Cal}} & 410   $\pm$20 & 26600  $\pm$1600 &18300  $\pm$3600 &150  $\pm$30 &6.8  $\pm$ 3.8 & 0.87$\pm$0.03\\ 
\textbf{{\color{cyan} ChaI}} & 150   $\pm$10 & 5380  $\pm$720 &640  $\pm$150 &25  $\pm$6 &1.1  $\pm$ 0.6  & 0.65$\pm$0.02   \\ 
\textbf{{\color{aquaMarine} ChaII-III}} & $"$& 3570  $\pm$480 &940  $\pm$220 &9  $\pm$2 &0.4  $\pm$ 0.2 & $"$ \\ 
\textbf{{\color{black} ChaEastI}} &$"$ & 1830 $\pm$250 &210  $\pm$50 &-  &- & $"$\\ 
\textbf{{\color{green} ChaEastII}} & $"$& 5870  $\pm$780 &350  $\pm$80 &- & -&  $"$ \\ 
\textbf{{\color{goldenRod} Musca}} & $"$& 5890  $\pm$590 &220  $\pm$50 &- & -& $"$ \\ 

\end{tabular}
\tablefoot{The SFR have been derived from the mass of dense molecular gas at \av >8 mag. according to the empirical relation expected from the “microphysics" of
prestellar core formation within filaments \citep{2017arXiv170500213S} : \mbox{SFR=$(4.5\pm2.5) \times 10^{-8}$ M$_{\odot}$ yr$^{-1}$ $\times$ (M$^{A_{\rm V}>8}_{\rm{H_2}} $/M$_{\odot}$).} }
\label{tab:CLmass}
\end{table*}

%% file: SubMass.txt
\renewcommand{\arraystretch}{1.25}
\setlength{\tabcolsep}{0.19cm}

\begin{table*}[p!]
\caption{Fractions of the total mass in the different gas phases for the 15 clouds and their CO-dark \hd mass fractions}
\centering
  \begin{tabular}{l|c |c |c |c||c|c}
Name & $F_{\rm{HI}}$ & $F_{\rm{DNM}}$ & $F_{\rm{CO}}$ & $F_{\rm{COsat}}$ &  $F_{\rm{COdk \, H_2}}^{50}$& $F_{\rm{COdk \, H_2}}^{100}$ \\ 
\hline 
\textbf{{\color{deepPink} Cet1}} & 0.79 $\pm$0.26 &0.17  $\pm$0.06 &0.03  $\pm$0.01 & - &0.73  $\pm$0.11 &0.84  $\pm$0.13   \\ 
\textbf{{\color{firebrick} Cet2}} & 0.79 $\pm$0.26 &0.13  $\pm$0.05 &0.07  $\pm$0.03 & - &0.50  $\pm$0.08 &0.67  $\pm$0.10   \\ 
\textbf{{\color{lime} tauM1}} & 0.69 $\pm$0.26 &0.19  $\pm$0.08 &0.10  $\pm$0.05 &0.010  $\pm$0.002 &0.49  $\pm$0.09 &0.66  $\pm$0.12   \\ 
\textbf{{\color{mediumOrchid} tauM2}} & 0.37 $\pm$0.13 &0.11  $\pm$0.05 &0.37  $\pm$0.16 &0.14  $\pm$0.05 & 0.11  $\pm$0.02 &0.19  $\pm$0.03   \\ 
\textbf{{\color{red} tauS1}} & 0.69 $\pm$0.14 &0.27  $\pm$0.06 &0.04  $\pm$0.01 & -&0.79  $\pm$0.04 &0.88  $\pm$0.05   \\ 
\textbf{{\color{darkGrey} tauS2}} & 0.71 $\pm$0.15 &0.25  $\pm$0.06 &0.03  $\pm$0.01 & - &0.82  $\pm$0.04 &0.90  $\pm$0.05   \\ 
\textbf{{\color{yellow} tauN1}} & 0.76 $\pm$0.38 &0.19  $\pm$0.11 &0.04  $\pm$0.03 &- & 0.68  $\pm$0.18 &0.81  $\pm$0.22   \\ 
\textbf{{\color{darkOrange} tauN2}} & 0.77 $\pm$0.39 &0.17  $\pm$0.10 &0.05  $\pm$0.03 &0.010  $\pm$0.005  &0.60  $\pm$0.16 &0.75  $\pm$0.20   \\ 
\textbf{{\color{dodgerBlue} Cal}} & 0.44 $\pm$0.09 &0.19  $\pm$0.04 &0.30  $\pm$0.09 &0.05  $\pm$0.01  &0.22  $\pm$0.01 &0.36  $\pm$0.02   \\ 
\textbf{{\color{navy} Auriga}} & 0.41 $\pm$0.12 &0.27  $\pm$0.09 &0.23  $\pm$0.09 &0.08  $\pm$0.02  &0.31  $\pm$0.04 &0.47  $\pm$0.06   \\ 
\textbf{{\color{goldenRod} Musca}} & 0.70 $\pm$0.18 &0.26  $\pm$0.07 &0.03  $\pm$0.01 &0.010  $\pm$0.002  &0.80  $\pm$0.08 &0.89  $\pm$0.09   \\ 
\textbf{{\color{cyan} ChaI}} & 0.68 $\pm$0.20 &0.20  $\pm$0.07 &0.08  $\pm$0.03 &0.04  $\pm$0.01 & 0.47  $\pm$0.06 &0.64 $\pm$0.09   \\ 
\textbf{{\color{aquaMarine} ChaII-III}} & 0.61 $\pm$0.18 &0.19  $\pm$0.07 &0.16  $\pm$0.06 &0.03  $\pm$0.01  &0.34  $\pm$0.05 &0.51  $\pm$0.07   \\ 
\textbf{{\color{black} ChaEastI}} & 0.85 $\pm$0.25 &0.05  $\pm$0.02 &0.10  $\pm$0.04 & -& 0.21  $\pm$0.03 &0.35  $\pm$0.05   \\ 
\textbf{{\color{green} ChaEastII}} & 0.81 $\pm$0.24 &0.13  $\pm$0.04 &0.05  $\pm$0.02 &- & 0.57  $\pm$0.08 &0.73  $\pm$0.10   \\ 

\end{tabular}

\tablefoot{$F_{\rm{COdk \, H_2}}^{50}$ and $F_{\rm{COdk \, H_2}}^{100}$ correspond to a 50$\%$ and 100$\%$ molecular DNM composition, respectively.
}
\label{tab:9mass}
\end{table*}